\documentstyle[aps,eqsecnum,epsf]{revtex}
\begin{document}

\title{Growth of nanostructures by cluster 
deposition : experiments and simple models}

\author{Pablo Jensen$^*$}

\address{D\'epartement de Physique des Mat\'eriaux, Universit\'e Claude
Bernard Lyon-1, 69622 Villeurbanne C\'edex, France}

\maketitle


\begin{abstract}
This paper presents a comprehensive analysis of simple models useful to 
analyze the growth of nanostructures obtained by cluster deposition. After
detailing the potential interest of nanostructures, I extensively
study the first stages of growth (the submonolayer regime) by kinetic
Monte-Carlo simulations. These simulations are performed in a wide variety 
of experimental situations : complete condensation, growth with reevaporation,
nucleation on defects, total or null cluster-cluster coalescence \ldots. 
The main scope of the paper is to help experimentalists
analyzing their data to deduce which of those processes are
important and to quantify them. A software including all these
simulation programs is available at no cost on request to the author. 
I carefully discuss experiments of growth from cluster beams and show how the 
mobility of the clusters on the surface can be measured : surprisingly high
values are found. An important 
issue for future technological applications of
cluster deposition is the relation between the size of the incident
clusters and the size of the islands obtained on the substrate. An
approximate formula which gives the ratio of the two sizes as a
function of the melting temperature of the material deposited is given. 
Finally, I study the atomic mechanisms which can explain the diffusion of 
the clusters on a substrate and the result of their mutual interaction 
(simple juxtaposition, partial or total coalescence \ldots).
\end{abstract}

\section{INTRODUCTION}

Growth of new materials with tailored properties is one of the most
active research directions for physicists. As pointed out by Silvan
Schweber in his brilliant analysis of the evolution of physics after
World War II : "An important transformation has taken place in
physics : As had previously happened in chemistry, an ever larger
fraction of the efforts in the field [are] being devoted to the study
of novelty [creation of new structures, new objects and new 
phenomena] rather than to the elucidation of fundamental laws and
interactions [...] Condensed matter physics has indeed become the study
of systems that have never before existed." \cite{schweber} 

Among these new materials, those presenting a structure controlled 
down to the nanometer scale are being extensively studied 
\cite{gleiter,palmernew,science,siegel_ency,cam,fendler,nanomat_revue}. 
There are different ways to build up nanostructured systems
 \cite{nanomat_revue} : atomic deposition \cite{atom_techno}, 
mechanical milling \cite{milling}, chemical methods \cite{fendler,shalaev}, 
gas-aggregation techniques \cite{siegel_ency,granqvist,LECD} \ldots 
Each of these techniques has its own advantages, but, as happens with atomic 
deposition techniques, the requisites of {\it control} 
(in terms of characterization and 
flexibility) and {\it efficiency} (in terms of quantity of matter
obtained per second) are generally incompatible. As a physicist wishing
to understand the details of the processes involved in the building 
of these nanostructures, I
will focus in this review on a carefully {\it controlled} method : low energy 
cluster deposition \cite{LECD}. Clusters are large "molecules" containing 
typically from 10 to 2000 atoms, and have been studied for their 
specific physical properties (mostly due to their large surface to volume 
ratio) which are {\it size dependent} and different from both the atoms and 
the bulk material \cite{science,deheer,joyes,microclusters,confs_clu}. 
By depositing {\it preformed}
clusters on a substrate, one can build nanostructures of two types :
in the submonolayer range, separated (and hopefully ordered) nanoislands, and
for higher thicknesses, thin films or cluster assembled materials (CAM).
The main advantage of the cluster deposition technique is that one can
carefully control the building block (i.e. the cluster) and characterize the
growth mechanisms. By changing the {\it size} of the
incident clusters one can change the growth mechanisms 
\cite{fuchs,brechignac} and the characteristics of the materials. 
For example, it has been shown that by changing
the mean size of the incident carbon clusters, one can modify the properties
of the carbon film, from graphitic to diamond-like \cite{paillard}.

This review is organized as follows.
First, I present briefly the interest of nanostructures, both in the
domain of nanoislands arranged on a substrate and as nanostructured, 
continuous films. I also review the different strategies employed
to deposit {\it clusters} on a substrate : by accelerating them
or by achieving their soft-landing. The scope of this section is to 
convince the reader that cluster deposition is a promising technique
for nanostructure growth in a variety of domains, and therefore deserves
a careful study. In Section \ref{sec_models}, models 
for cluster deposition are introduced. These models can also be useful 
for {\it atomic} deposition in some simple cases, namely when aggregation
is irreversible. The models are adapted here to the physics of {\it cluster} 
deposition. In this case, reevaporation
from the substrate can be important (as opposed to the usual conditions
of Molecular Beam Epitaxy), cluster-cluster aggregation is always irreversible 
(as opposed to the possibility of bond breaking for atoms \cite{reversible})
 and particle-particle {\it coalescence} is possible. After a 
brief presentation of Kinetic Monte-Carlo
(KMC) simulations, I show how the submonolayer regime can be
studied in a wide variety of experimental
situations : complete condensation, growth with reevaporation,
nucleation on defects, formation of two and three dimensional islands \ldots
Since I want these models to be useful for experimentalists, Section
\ref{sec_interpret} is entirely devoted to the presentation of a strategy
on how to analyze experimental data and extract microscopic parameters
such as diffusion and evaporation rates. I remind the reader that
a simple software simulating all these situations is available at no 
cost on simple request to the author. Section \ref{exps}
analyzes in detail several experiments of cluster deposition. These studies
serve as examples of the recipes given in Section \ref{sec_interpret} 
to analyze the data and also to demonstrate that clusters can have 
surprisingly large mobilities (comparable to atomic mobilities) on 
some substrates. A first interpretation of these intriguing results  
at the atomic level is given in Section \ref{sec_atomic}, where the
kinetics of cluster-cluster coalescence is also studied. The main
results of this Section are that 
high cluster mobilities can be achieved provided the cluster does not find
an epitaxial arrangement on the substrate and that cluster-cluster
coalescence can be much slower than predicted by macroscopic theories.

A note on terminology : The structures 
formed on the surface by aggregation of the clusters are called
islands. This is to avoid the possible
confusion with the terms usually employed for atomic deposition where
the clusters are the islands formed by aggregation of atoms on the
surface. Here, the clusters are {\it preformed} in the gas phase {\it
before} deposition. I use {\it coverage} for the actual portion of the
surface covered by the islands and {\it thickness} for the total
amount of matter deposited on the surface (see also Table I).

\section{INTEREST AND BUILDING OF NANOSTRUCTURES}

Before turning to the heart of this paper - the growth of nanostructures
by cluster deposition - I think it is appropriate to show why one wants
to obtain nanostructures at all and how these can be prepared experimentally.
Due to the technological impetus, a tremendous amount of both experimental
and theoretical work has been carried out in this field, and it is
impossible to summarize every aspect of it here. For a recent and rather
thorough review, see Ref. \cite{nanomat_revue} where the possible technological
impact of nanostructures is also addressed. Actually, several journals are
entirely devoted to this field \cite{revues}. The reader is also referred to 
the enormous number of World Wide Web pages (about 6000 on nanostructures),
especially those quoted in Ref. \cite{www_nano}. A short 
summary of the industrial interest of nanostructures \cite{eur_news} and 
introductory reviews on the interest of "Nanoscale and ultrafast devices" 
\cite{tod90} or "Optics of nanostructures" \cite{tod93} have appeared recently. 

There are two distinct (though related) domains where nanostructures can
be interesting for applications. The first stems from the desire of
miniaturization of electronic devices. Specifically, one would like to grow
organized nanometer size islands with specific electronic properties. As a
consequence, an impressive quantity of deposition techniques have 
been developed to grow carefully controlled thin films and nanostructures from
atomic deposition \cite{atom_techno}. While most of these techniques are 
complex and keyed to specific applications, Molecular Beam Epitaxy (MBE) 
\cite{mbe} has received much attention from the physicists \cite{lagally}, 
mainly because of its (relative) simplicity. The second subfield is that 
of nanostructured materials \cite{nanomat_revue}, as thin or thick films, 
which show (mechanical, catalytic, optical) 
properties different from their microcrystalline counterparts
\cite{siegel_ency,cam,shalaev,camnato,jena}.

I will now briefly review the two subfields since cluster deposition can 
be used to build both types of nanostructures. Moreover, some of the 
physical processes studied below (such as cluster-cluster coalescence)
are of interest for both types of structure.

\subsection{Organized nanoislands}

There has been a growing interest for the fabrication of organized 
islands of nanometer dimensions. One of the reasons
 is the obvious advantage of miniaturizing the electronic devices both
for device speed and density on a chip (for a simple 
and enjoyable introduction to the progressive miniaturization of electronics 
devices, see Ref. \cite{turton}). But it should be noted that at these
scales, shrinking the size of the devices does also change their
properties, owing to quantun confinement effects. Specifically, 
semiconductor islands smaller than the Bohr diameter of the bulk material (from
several nm to several tens of nm) show interesting properties : as their
size decreases, their effective bandgap increases. The possiblity to
tailor the electronic properties of a given material by playing on its
size has generated a high level of interest in the field of these {\it 
quantum dots} \cite{qd}. But quantum dots are not the only driving force 
for obtaining organized nanoislands. Isolated nanoparticles are also 
interesting as model catalysts (see Refs. \cite{model_catalysis,revhenry}
and Chapter 12 of Ref. \cite{nanomat_revue}). Clearly,
using small particles increases the specific catalytic area for a given
volume. More interesting, particles 
smaller than 4-5 nm in diameter might show specific catalytic properties,
different from the bulk \cite{revhenry,che}, although the precise mechanisms
are not always well identified (Chapter 12 of Ref. \cite{nanomat_revue}). One
possibility is the increase, for small particle sizes, of the proportion
of low coordination atoms (corners, kinks) whose electronic (and therefore
catalytic) properties are
expected to be different from bulk atoms. For even smaller particles 
(1-2 nm), the interaction with the substrate can significantly alter
their electronic properties \cite{intercata}. 
Recently, there have been attempts at
{\it organizing} the isolated islands to test the consequences on the
catalytic properties \cite{organized_cata}. Obtaining isolated clusters on a 
surface can also be interesting to study their
properties. For example, Schaefer et al. \cite{schaefer} have obtained 
isolated gold clusters onto a variety of substrates to investigate the elastic
properties of {\it single} nanoparticles by Atomic Force Microscopy (AFM).

Let me now briefly turn on to the the possible ways of obtaining such
organized nanoislands. Deposition of atoms on carefully controlled substrates 
is the main technique 
used presently by physicists to try to obtain a periodic array of nanometer islands of well-defined sizes. 
A striking example \cite{revharri} of organized nanoislands is given 
in Fig. \ref{triangles}.
These triangular islands have been grown on the dislocation network formed
by the second Ag atomic layer on Pt(111). Beautiful as these triangles are,
they have to be formed by nucleation and growth
on the substrate, and therefore the process is highly dependent on the
interaction of the adatoms with the substrate (energy barriers for diffusion, 
possibility of exchange of adatoms and substrate atoms, \ldots). This
drastically limits the range of possible materials that can be grown
by this method. However, the growth of strained islands 
by heteroepitaxy is under active study, since stress is a force 
which can lead to order, and even a tunable order, as observed for example
in the system $PbSe/Pb_{1-x}Eu_xTe$ \cite{springholz}  
(see Refs. \cite{spontaneous_qd,elastic_exp} for further details on stress).

\vspace{1cm}

In this review, I will focus on an alternative approach to form nanoislands
on substrates : instead of growing them by atom-atom aggregation {\it on} 
the substrate, a process which dramatically depends
on the idiosyncrasies of the substrate and its interaction with the
deposited atoms, one can prepare the islands (as free clusters) {\it before}
deposition and then deposit them. It should be
noted that the cluster structure can be extensively characterized
{\it prior to} deposition by several in-flight techniques such as 
time-of-flight spectrometry, photo-ionization or 
fragmentation \cite{vialle}. Moreover, the properties of these
building blocks can be adjusted by changing their {\it size}, which
also affects the growth mechanisms, and therefore the film
morphology \cite{fuchs,brechignac}. A clear example of the possibility to
change the film morphology by varying only the mean cluster size has been
given a few years ago by Fuchs et al. \cite{fuchs} (Fig. \ref{sbfuchs})
and this study has been completed recently by Brechignac's group for 
larger cluster sizes \cite{brechignac}. There are several
additional interests for depositing clusters. First, these are grown 
in extreme nonequilibrium 
conditions, which allows to obtain metastable structures or alloys. It 
is true that neither islands grown on a 
substrate are generally in equilibrium, but the quenching rate is very 
high in a beam, and the method is more flexible since one avoids the effects
of nucleation and growth on a specific substrate. For example, PdPt alloy
clusters - which are known to have interesting catalytic properties - can be prepared with a precise composition (corresponding to the
composition of the target rod, see below) and variable size and then
deposited on a surface \cite{ptpd}. The same is true for SiC clusters
where one can modify the electronic properties of the famous $C_{60}$
clusters by introducing in a controlled way Si atoms before deposition
\cite{sic}. This allows to tune within a certain range the properties of 
the films by choosing the preparation conditions of the preformed clusters.
It might also be anticipated that cluster nucleation is less sensitive to 
impurities than atomic nucleation. Atomic island growth can be 
dramatically affected by them, as exemplified by the celebrated case 
of the different morphologies of Pt islands grown on Pt(111) \cite{comsa} 
which were actually the result of CO contamination at an
incredibly low level : $10^{-10}$ mbar \cite{triangles_impu}. Instead,
clusters, being larger entities, might interact less specifically with the 
substrate and its impurities. There is still no systematic way of organizing 
the clusters on a surface.
One could try to pin them on selected sites such as defects or to 
encapsulate the clusters 
with organic molecules before deposition in order to obtain ordered arrays 
on a substrate \cite{andres}. 

\subsection{Nanostructured materials}

Although my main focus in this review is the understanding of the 
first stages of growth, it is worth pointing out the interest of
thicker nanostructured films (for a recent review of this field, see Ref. 
\cite{nanomat_revue}). It is known \cite{siegel_ency,cam,camnato} that the
(magnetic, optical and mechanical) properties of these films can 
be intrinsically different from their macrocrystalline counterparts. 
The precise reasons for this are currently being investigated, but one 
can cite the presence of a significant fraction (more than 10 \%) of atoms 
in configurations different from the
bulk configuration, for example in grain boundaries \cite{gleiter}. It
is reasonable to suppose that both dislocation generation and
mobility may become significantly difficult in nanostructured films
\cite{siegel_ency}. For example, recent studies of the mechanical 
deformation properties of nanocrystalline copper \cite{schiotz} have shown 
that high strain can be reached before the appearance of plastic 
deformation. A review of the effects of nanostructuration on the
mechanical response of solids is given by Weertman and Averback in 
chapter 13 of Ref. \cite{nanomat_revue}.
Another interesting property of these materials is that their crystalline
order is intermediate between that of the amorphous 
materials (first neighbor order) and of crystalline materials (long range 
order). It is given by the size of the crystalline cluster, which can be tuned.
For example, for random magnetic materials, by varying the size of the
clusters, and consequently of the ferromagnetic domain, one can study the
models of amorphous magnetic solids \cite{magnetic}.

\subsection{How can one deposit clusters on surfaces?}

After detailing the potential interests of nanostructures, I now
address the practical preparation methods by {\it cluster} deposition.
Two main variants have been explored. Historically, the
first idea has been to produce beams of {\it accelerated} (ionized)
clusters and take advantage of the incident kinetic energy to enhance
atomic mobility even at low substrate temperatures. This method
does not lead in general to nanostructured materials, but to films
similar to those obtained by atomic deposition, with sometimes better
properties. A more recent approach is
to deposit neutral clusters, with {\it low energy} to preserve their
peculiar properties when they reach the surface. The limit between the
two methods is roughly at a kinetic energy of 0.1 to 1 eV/atom.  

\subsubsection{Accelerated clusters}

The Japanese group of Kyoto University was the first to explore the idea
of depositing clusters with high kinetic energies (typically a few keV)
to form thin films \cite{yamada}. The basic idea of the Ionized Cluster
Beam (ICB) technique is that the cluster breaks upon
arrival and its kinetic energy is transferred to the adatoms 
which then have high lateral (i.e. parallel to the substrate) 
mobilities on the surface. This allows in
principle to achieve epitaxy at low substrate temperatures, which is
interesting to avoid diffusion at interfaces or other activated
processes. Several examples of good epitaxy by ICB have been obtained by
Kyoto's group : Al/Si \cite{alsi} which has a large mismatch and many other
couples of metals and ceramics on various crystalline substrates such
as Si(100), Si(111) \ldots Molecular Dynamics (MD) simulations have supported
this idea of epitaxy by cluster spreading \cite{biswas}. 
The reader is referred to Yamada's reviews
\cite{yamada} for an exhaustive list of ICB applications, which also 
includes high energy 
density bombardment of surfaces to achieve sputter yields significantly
higher than obtained from atomic bombardment \cite{insepov}.

However, the physics behind these technological successes is not clear.
In fact, the very presence of a significant fraction of large clusters 
in the beam seems dubious \cite{turner,jarold}. There is some 
experimental evidence \cite{yamada} offered by Kyoto's group to 
support the effective presence 
of a significant fraction of large clusters in the beam, but the 
evidence is not conclusive. In short, it is difficult to make a 
definite judgement about the ICB technique.
There is no clear proof of the presence of clusters in the beam and
the high energy of the incident particles renders difficult any attempt
of modelling. Kyoto's group has clearly shown that ICB does lead to 
good quality films in many cases but it is not
clear how systematic the improvement is when compared to atomic deposition techniques.

Haberland's group
in Freiburg has developed recently a different technique called
Energetic Cluster Impact (ECI) where a better controlled beam of
energetic clusters is deposited on surfaces \cite{eci_exp}. Freiburg's
group has shown that accelerating the clusters leads to improvements
in some properties of the films :
depositing slow clusters (energy per atom 0.1 eV) produces metal
films which can be wiped off easily, but accelerating them before deposition
(up to 10 eV per atom) results in strongly adhering films \cite{eci_md}.
MD simulations of cluster deposition \cite{eci_md} have explained 
qualitatively this behavior : while low energetic clusters tend to pile up
on the substrate leaving large cavities, energetic clusters lead to a compact 
film (Fig. \ \ref{md_eci}). It is interesting to note that, even for the
highest energies explored in the MD simulations (10 eV per atom), no atoms
were ejected form the cluster upon impact. The effect of film smoothening
is only due to the flattening of the cluster when it touches the substrate. 
Some caution on the interpretation of
these simulations in needed because of the very short time scales which
can be simulated (some ps). Similar MD simulations of the impact of a cluster
with a surface at higher energies have also been performed \cite{massobrio}.
Recently, Palmer's group \cite{palmerprl} has studied the interaction
of Ag clusters on graphite for various incident kinetic energies (between
15 and 1500 eV). They have shown that, for {\it small} ($Ag_3$) clusters, 
the probability of a cluster penetrating the substrate or not critically 
depends on its orientation relative to the substrate. 

\subsubsection{Low energy clusters}

Another strategy to grow nanostructures with cluster beams consists in
depositing {\it low energy} particles 
\cite{LECD,brechignac,andres,palmer,palmerdepo,wang,ma,divers_depo_clu}. 
Ideally, by depositing the clusters with low kinetic
energies, one would like to conserve the memory of the free cluster
phase \cite{LECD} to form thin films with original properties. Since
the kinetic energy is of the order of 10 eV per cluster \cite{roux}, i.e. a few
meV per atom which is negligible compared to the binding energy of an
atom in the cluster, no fragmentation of the clusters is
expected upon impact on the substrate. 
Fig. \ \ref{md_eci} suggests that the films are 
porous \cite{eci_md,LECDnano}, which is interesting to keep one of 
the peculiarities of the clusters : their high surface/volume ratio which 
affects all the physical (structural,
electronic) properties as well as the chemical reactivity (catalysis).
Concerning deposition of carbon clusters, experiments \cite{LECD,paillard} 
as well as simulations \cite{galli} have shown that the carbon clusters
preserve their identity in the thick film. Another interesting type of
nanostructured film grown by cluster deposition is
the growth of cermets by combining a cluster beam with an atomic beam
of the encapsulating material \cite{cermet}.  The point is that the
size of the metallic particles is determined by the incident cluster
size and the concentration by the ratio of the two fluxes. Then, these
two crucial parameters can be varied independently, in contrast to the
cermets grown from atomic beams and precipitation upon annealing.

Cluster beams are generated by different techniques : Multiple Expansion
Cluster Source (MECS, \cite{schaefer}), gas-aggregation \cite{granqvist,LECD,palmer,sattler,rayane} \ldots All these techniques 
produce a beam of clusters with a distribution 
of sizes, with a dispersion of about half the mean size. For simplicity,
I will always refer to this mean size. In gas-aggregation techniques,
 an atomic vapor obtained from
a heated crucible is mixed with an inert gas (usually Ar or He) and the 
two are cooled by adiabatic expansion, resulting in supersaturation
and cluster formation. The mean cluster size can be monitored by the
different source parameters (such as the inert gas pressure) and can be 
measured by a time of flight mass spectrometer. For further experimental 
details on this technique, see Refs. \cite{LECD,elshall}. To produce clusters 
of refractory materials, a different evaporation technique is needed :
laser vaporization \cite{LECD,elshall,laser}.  A plasma
created by the impact of a laser beam focused on a rod of is thermalized 
by injection of a high pressure
He-pulse (typically, 3-5 bars during 150 to 300 $\mu$s), which permits the
cluster growth. The mean cluster size is governed by several
parameters such as the helium flow, the laser power and the delay time
between the laser shot and the helium pulse. As a consequence of the
pulsed laser shot, the cluster flux reaching the surface is not
continuous but {\it chopped}.
Typical values for the chopping parameters are : active portion of the
period $\simeq 100\mu s$ and chopping frequency $f=10Hz$.

\subsubsection{Other approaches}

Alternatively, one can deposit accelerated clusters onto a buffer layer 
which acts as a "mattress" to dissipate the kinetic energy. This layer 
is then evaporated,
which leads to cluster soft-landing onto the substrate \cite{lausanne,landman}.
The advantage of this method is that it is possible to select the mass
of the ionized clusters before deposition. However, it is difficult
with this technique to reach high enough deposition rates to grow films
in reasonable times. Vitomirov
et al \cite{vitomirov} deposited {\it atoms} onto a rare-gas buffer layer :
the atoms first clustered on top and within the layer which was afterwards
evaporated, allowing the clusters to reach the substrate.
Finally, deposition of clusters from a
Scanning Tunneling Microscope (STM) tip has been shown to be possible, 
both theoretically \cite{tiptheo} and experimentally \cite{tipexp}.

\section{MODELS OF PARTICLE DEPOSITION}
\label{sec_models}

I describe in this section simple models which allow to understand
the first stages of film growth by low energy cluster deposition. 
These models can also be useful for understanding 
the growth of islands from {\it atomic} beams in the submonolayer regime in 
{\it simple} cases, namely (almost) perfect substrates, irreversible 
aggregation, etc. and they have allowed 
to understand and quantify many aspects of the growth : for a 
review of analysis of atomic deposition with this kind of models, see 
Refs. \cite{lagally,revharri,zhang,larecherche}. The models
described below are similar to previous models of diffusing particles 
that aggregate, but such
``cluster-cluster aggregation'' (CCA) models \cite{meakin} do not
incorporate the possibility of continual injection of new particles via
deposition, an essential ingredient for thin film growth.

\vspace{.4cm}

Given an experimental system (substrate and cluster chemical nature), 
how can one predict the growth characteristics for a given set of 
parameters (substrate temperature, incoming flux of clusters \ldots)?

A first idea - the "brute-force" approach - would be to run a
Molecular Dynamics (MD) simulation with ab-initio potentials for the particular
system one wants to study. It should be clear that such an approach is bound 
to fail since the calculation time is far too large for present-day
computers. Even using empirical potentials (such as Lennard-Jones, Embedded
Atom or Tight-Binding) will not do because 
there is an intrinsically large time scale in the growth problem : the 
mean time needed to fill a significant fraction of the substrate with the
incident particles. An estimate of this time is fixed by $t_{ML}$, the time 
needed to fill a monolayer : $t_{ML} \simeq 1/F$ where $F$ is the particle 
flux expressed in monolayers per second (ML/s). Typically, the experimental
values of the flux are lower than 1ML/s, leading to $t_{ML} \geq 1s$.
Therefore, there is a time span of about 13 decades between the
typical vibration time ($10^{-13}s$, the lower time scale for the 
simulations) and $t_{ML}$, rendering hopeless any "brute-force" approach.

There is a rigorous way \cite{voter} of circumventing this time span problem : 
the idea is to "coarsen" the description by defining {\it elementary 
processes}, an approach somewhat reminiscent of the usual (length, energy) 
renormalization of particle physics \cite{schweber}. One "sums up" all
the short time processes (typically, atomic thermal vibrations) in effective
parameters (transition rates) valid for a higher level (longer time)
description. I will now briefly
describe this rigorous approach and then proceed to show how it can be adapted
to cluster deposition.

\subsection{Choosing the elementary processes}

Voter \cite{voter} showed that the interatomic
potential for any system can be translated into a finite set of parameters,
which then provides the exact dynamic evolution 
of the system. Recently, the same idea has been applied to Lennard-Jones
potentials \cite{lj2p} by using only {\it two} parameters. The point is that 
these coarse-grained, lattice-gas approach needs orders of magnitude 
less computer power than the MD dynamics described above. One can understand
the basic idea by the following simple example : for the MD description of
the diffusion of an atom by hopping, one has to follow in detail its motion 
at the picosecond
scale, where the atom mainly oscillates in the bottom of its potential well.
Only rarely at this time scale will the atom jump from site to site, which
is what one is interested on. Voter showed that, provided some
conditions are met concerning the separation of these two
time scales, and restricting the motion to a regular (discrete) lattice
(see \cite{voter} for more details), one could replace this
"useless" information by an effective parameter taking into account all
the detailed motion of the atom within the well (including the correlations
between the motions of the atom and its neighbors) and allowing a rapid
evaluation of its diffusion rate.

Unfortunately, this rigorous approach is not useful for 
{\it cluster} deposition, because the number of atomic degrees of freedom
(configurations) is too high. Instead, one chooses - from physical
intuition - a "reasonable" set of elementary processes, whose magnitudes
are used as {\it free parameters}. This allows to understand the role of
each of these elementary processes during the growth and then to fit
their value from experiments (Fig. \ref{kmccategories}). These are the 
models which I will study in this paper, with precise examples
of parameter  fit (see section \ref{exps}). 
Examples of such fits from experimental data for
{\it atomic} deposition include homoepitaxial
growth of GaAs(001) \cite{shitara}, of Pt(100) \cite{linderoth} or 
several metal(100) surfaces \cite{bartelt_vac}. Of course, fitting is
not very reliable when there are too many (almost free) parameters. An
interesting alternative are intermediate cases, where parameters are 
determined from known potentials
but with a simplified fitting procedure taking into account what is
known experimentally of the system under study : see Ref. \cite{ferrando} 
for a clear example of such a possibility.

\subsection{Predicting the growth from the selected elementary processes}

To be able to adjust the values of the elementary processes from
experiments, one must first predict the growth from these processes. 
The oldest way is to write "rate-equations" which
describe in a mean-field way the effect of these processes on the
number of isolated particles moving on the substrate (called monomers) 
and islands of a
given size. The first author to attempt such an approach for growth
was Zinsmeister \cite{zinsmeister} in 1966, but the general 
approach is similar to the rate-equations first used by 
Smoluchovsky for particle aggregation \cite{smoluchovsky}. In
the seventies, many papers dealing with better mean-field
approximations and applications of these equations to interpret
experimental systems were published. The reader is referred to
the classical reviews by Venables and co-workers 
\cite{venables73,venables84} and Stoyanov and Kaschiev \cite{stoyanov}
for more details on this approach. More recently, there have
been two interesting improvements. The first is by Villain and
co-workers which have simplified enormously the mathematical
treatment of the rate-equations, allowing one to understand
easily the results obtained in a variety of cases \cite{villain,vipi}.
Pimpinelli et al. have recently published a summary of the application
of this simplified treatment to many practical situations using a
unified approach \cite{pimpi}. The second improvement is due to Bales 
and Chrzan \cite{bales} who
have developed a more sophisticated self-consistent rate-equations approach
which gives better results and allows to justify many of the 
approximations made in the past. However, these analytical approaches
are mean-field in nature and cannot reproduce all the characteristics
of the growth. Two known examples are the island morphology and the
island size distribution \cite{bales}.

The alternative approach to predict the growth are  
Kinetic Monte-Carlo (KMC) simulations. KMC simulations
are an extension of the usual Monte Carlo 
\cite{metropolis,kmc_springer,kmc_base} algorithm and
provide a rigorous way of calculating the dynamical evolution of a complicated
system where a large but {\it finite} number of random processes 
occur at given rates. 
KMC simulations are useful when one chooses to deal with only the slowest
degrees of freedom of a system, these variables being only weakly
coupled to the fast ones, which act as a heat bath \cite{kmc_base}. 
The "coarsened" description of film growth (basically, diffusion) given above
is a good example \cite{voter,bales,tang,model,boston,laszlo}, but other 
applications \cite{kmc_base} of KMC 
simulations include interdiffusion in alloys, slow phase separations \ldots 
The principle of KMC simulations is straightforward :
one uses a list of all the possible processes together with their
respective rates $\nu_{pro}$ and generates the time evolution of the system
from these processes taking into account the random character of the evolution.
For the simple models of film growth described below, systems
containing up to 4000 x 4000 lattice sites can be simulated in
a reasonable time (a few hours), which limits the finite size effects 
usually observed in this kind of simulation. Let me now
discuss in some detail the way KMC simulations are implemented to reproduce
the growth, once a set of processes has been defined, with their
respective $\nu_{pro}$ taking arbitrary values or being derived from 
known potentials.

There are two main points to discuss here : the physical 
correctness of the dynamics and the calculation speed. Concerning the
first point, it should be noted that, originally \cite{metropolis}, 
Monte Carlo simulations aimed at the description of
the {\it equations of state} of a system. Then, "the MC method performs a 
"time" averaging of a model with (often artificial) stochastic kinetics [...] :
time plays the role of a label characterizing the sequential order of
states, and need not be related to the physical times" \cite{binder}.
One should be cautious therefore on the precise Monte Carlo 
scheme used for the simulation when attempting at describing the kinetics 
of a system, as in KMC simulations. For example, there are doubts \cite{metiudiff,kang} about some simulation work \cite{wrongkinetics,khare} 
carried out using Kawasaki dynamics. This point is discussed in great detail 
in Ref. \cite{kang}. 

Let me address now the important problem of the calculation speed. One 
could naively think of choosing a time interval $\Delta t$
smaller than all the relevant times in the problem, and then repeat
the following procedure : 

(1) choose one particle randomly   

(2) choose randomly one of the possible processes for this particle

(3) calculate the probability $p_{pro}$ of this process happening 
during the time interval $\Delta t$ ($p_{pro}=\nu_{pro} \Delta t $)

(4) throw a random number $p_r$ and compare it with $p_{pro}$ : if 
$p_{pro} < p_r$ perform the process, if not go to the next step 

(5) increase the time by $\Delta t$ and goto (1)

This procedure leads to the correct kinetic evolution of the system
but might be extremely slow if there is a large range of probabilities
$p_{pro}$ for the different processes (and therefore some $p_{pro} \ll 1$). 
The reason is that a significant fraction of the loops leads to rejected 
moves, i.e. to no evolution at all of the system.

Instead, Bortz et al. \cite{bortz} have proposed a clever approach
to eliminate {\it all} the rejected moves and thus reduce dramatically
the computational times. The point is to choose not the particles but
the {\it processes}, according to their respective rate and the number of
possible ways of performing this process (called $\Omega_{pro}$). This 
procedure can be represented schematically as follows :

(1) update the list of the possible ways of performing the processes $\Omega_{pro}$

(2) randomly select one of the process, {\it weighting} the probability
of selection by the process rate  $\nu_{pro}$ and $\Omega_{pro}$ : 
$p_{pro} = (\nu_{pro} \Omega_{pro})/\left({\sum_{processes}^{}{\Omega_{pro} \nu_{pro}}}\right)$

(3) randomly select a particle for performing this process

(4) move the particle 

(5) increase the time by $dt=\left({\sum_{processes}^{}{\Omega_{pro} \nu_{pro}}}\right)^{-1}$ 

(6) goto (1)

A specific example of such a scheme for cluster deposition
is given below (Section \ref{basic_clu}). Note that the new procedure 
implies a less intuitive increment of time, and that one has
to create (and update) a list of all the $\Omega_{pro}$ constantly, 
but the acceleration of the calculations is worth the effort.

A serious limitation of KMC approaches is that one has to assume a
finite number of local environments (to obtain a finite number of
parameters) : this confines KMC approaches to regular lattices, thus
preventing a rigorous consideration of elastic relaxation, stress effects \ldots
everything that affects not only the {\it number} of first or second nearest
neighbors but also their precise position. Indeed, considering the 
precise position as in MD simulations introduces a {\it continuous} variable
and leads to an infinite number of possible configurations or processes. 
Stress effects can be introduced
approximately in KMC simulations, for example  \cite{ratsch} by allowing 
a variation of the bonding energy of an atom to an island as a function of the
island size (the stress depending on the size), but it is unclear how
meaningful these approaches are (see also Refs. 
\cite{diff_strain}). I should quote here a recent proposition \cite{hamfk}
inspired on the old Frenkel-Kontorova model \cite{fk} which allows to 
incorporate some misfit effects in rapid simulations. It remains to
explore whether such an approach could be adapted to the KMC scheme.

\subsection{Basic elementary processes for cluster growth}
\label{basic_clu}

What is likely to occur when clusters are deposited on a surface ? I will
present here the elementary processes which will be used in cluster
deposition models : deposition,
diffusion and evaporation of the clusters and their interaction on the
surface (Figs. \ \ref{dda} and \ref{interaction}). The influence of
surface defects which could act as traps for the particles is also
addressed.

A simple physical rationale for choosing only a limited set of
parameters is the following (see Fig. \ \ref{timescale}). For any given system,
there will be a "hierarchy" of time scales, and the relevant ones
for a growth experiment are those much lower than $t_{ML} \simeq 1/F$.
The others are too slow to act and can be neglected. The hierarchy of
time scales (and therefore the relevant processes) depend of course on the
precise system under study. It should be noted that for {\it cluster} 
deposition the situation is somewhat simpler than for atom deposition 
\cite{crit_rev} since many elementary
processes are very slow. For example, diffusion of clusters on top of
an already formed island is very low \cite{ss}, cluster detachment
from the islands is insignificant and edge diffusion is not an elementary
process at all since the cluster cannot move as an entity over the island
edge (as I will discuss in section \ref{sec_coale}, the equivalent process
is cluster-cluster coalescence by atomic motion). Let me now discuss in detail
each of the elementary processes useful for cluster deposition.

The first ingredient of the growth, {\it deposition}, is quantified by 
the flux $F$, i.e. the number of clusters that are
deposited on the surface per unit area and unit time. The flux is
usually uniform in time, but in some experimental situations it can be
pulsed, i.e. change from a constant value to 0 over a given period. 
Chopping the flux can affect the growth of
the film significantly \cite{hache}, and I will take this into
account when needed (Section \ref{fhach}).

The second ingredient is the {\it diffusion} of the clusters which
have reached the substrate. I assume that the diffusion is brownian,
i.e. the particle undergoes a random walk on the substrate. To quantify
the diffusion, one can use both the usual diffusion coefficient $D$ or
the diffusion time $\tau$, i.e. the time needed by a cluster to move
by one diameter. These two quantities are connected by $D = d^2/(4\tau)$ 
where $d$ is the diameter of the cluster. Experiments
show that the diffusion coefficient of a cluster can be surprisingly
large, comparable to the atomic diffusion coefficients. The diffusion
is here supposed to occur on a perfect substrate. Real surfaces always
present some defects such as steps, vacancies or adsorbed chemical
impurities. The presence of these defects on the surface could
significantly alter the diffusion of the particles and therefore the
growth of the film. I will include here one simple kind of defect, a
perfect trap for the clusters which definitively prevents them from
moving.

A third process which could be present in growth is {\it
re-evaporation} of the clusters from the substrate after a time
$\tau_e$. It is useful to define $X_S=\sqrt{D\tau_e}$ the mean
diffusion length on the substrate before desorption.

The last simple process I will consider is the {\it interaction}
between the clusters.  The simplest case is when aggregation is
irreversible and particles simply remain juxtaposed upon contact.
This occurs at low temperatures. At higher temperatures,  cluster-cluster 
coalescence will be active (Fig. \ \ref{interaction}). 
Thermodynamics teaches us that
coalescence should always happen but without specifying the kinetics.
Since many clusters are deposited on the surface per unit time,
kinetics is here crucial to determine the shape of the islands formed
on the substrate. A total comprehension of the kinetics is still
lacking, for reasons that I will discuss later (Section \ref{sec_coale}). 
I note that the shape of
the clusters and the islands on the surface need not be perfectly
spherical, even in the case of total coalescence. Their interaction
with the substrate can lead to half spheres or even flatter shapes
depending on the contact angle. Contrary to what happens for
atomic deposition, a cluster touching an
island forms a huge number of atom-atom bonds and will not detach from it.
Thus, models including {\it reversible} particle-particle aggregation 
\cite{reversible} are not useful for cluster deposition.

The specific procedure to perform a rapid KMC simulation of a system (linear
size L) when deposition, diffusion and evaporation of the monomers are
 included is the following. The processes are : deposition of a particle 
($\nu_{depo}=F$, $\Omega_{depo}=L^2$ (it is possible to deposit a particle 
on each site of the lattice)), diffusion of a monomer ($\nu_{diff}=1/\tau$, 
$\Omega_{diff}= \rho L^2$ where $\rho$ is the monomer
density on the surface) and evaporation of a monomer 
($\nu_{evap}=1/\tau_e$, $\Omega_{evap}=\rho L^2$). For each loop, one
calculates two quantities $p_{drop} = F / (F + \rho ({1\over \tau_e} +
{1\over \tau}))$ and $p_{dif} = (\rho /\tau) / (F + \rho ({1\over
\tau_e} + {1\over \tau}))$. Then,
one throws a random number p ($0 < p < 1$) and compare it to $p_{drop}$
and $p_{dif}$.  If $p < p_{drop}$, a particle is deposited in a random
position; if $p > p_{drop} + p_{dif}$, a monomer (randomly selected) is removed,
otherwise we just move a randomly chosen monomer.  
After each of these possibilities, one checks
whether an aggregation has taken place (which modifies the
number of monomers on the surface, and therefore the number of possible 
diffusion or evaporation moves), increase the time by 
$dt=1/(F L^2 + \rho L^2 ({1\over \tau_e} + {1\over \tau}))$ and go to the 
next loop.

\vspace{.7cm}

The usual game for theoreticians is to combine these elementary
processes and predict the growth of the film.  However,
experimentalists are interested in the reverse strategy : from (a set
of) experimental results, they wish to understand which elementary
processes are actually present in their growth experiments and what
are the magnitudes of each of them, what physicists call understanding
a phenomenon. The problem, of course, is that with so many processes,
many combinations will reproduce the same experiments (see specific
examples below). Then, some clever guesses are needed to first
identify which processes are present. For example, if the saturation
island density does not change when flux (or substrate temperature) is
changed, one can guess that nucleation is mostly occurring on defects of
the surface.

In view of these difficulties, next
section is devoted to predict the growth when the microscopic
processes (and their values) are known. After, in Section \ref{sec_interpret}, 
I propose a detailed procedure to identify and quantify the microscopic 
process from the experiments. Finally, Section \ref{exps} reviews the
experimental results obtained for cluster deposition.

\section{PREDICTING GROWTH WITH COMPUTER SIMULATIONS}
\label{simu}

The scope of this section is to find formulas or graphs to deduce the
values of the microscopic processes (diffusion, evaporation \ldots )
from the observed experimental quantities (island density, island size
histograms \ldots ). The "classical"
studies \cite{venables73,venables84,stoyanov} have focused on the 
evolution of the
concentration of islands on the surface as a function of time, and
especially on the saturation island density, i.e. the maximum of the
island density observed before reaching a continuous film. The reason
is of course the double possibility to calculate it from
rate-equations and to measure it experimentally by conventional
microscopy. I will show other interesting quantities such as island
size distributions which are measurable experimentally and have been
recently calculated by computer simulations 
\cite{evaprb,mrs,smilauer,japan,stroscio,evans}. 

I will study the two limiting cases of pure juxtaposition and total coalescence 
(which are similar to two and three dimensional growth in atomic 
deposition terminology) separately. Experimentally, the distinction between the
two cases can be made by looking at the shape of the 
supported islands : if they are
circular (and larger than the incident clusters) they have been formed
by total coalescence; if ramified by pure juxtaposition (see several
examples below, section \ref{exps}).

In both cases, I analyze how the
growth proceeds when different processes are at work : diffusion,
evaporation, defects acting as traps, island mobility \ldots  In the
simulations, I often take the diffusion time $\tau$ to be the unit
time : in this case, the flux is equivalent to the normalized flux $\phi$ 
(see the Table) and
the evaporation time corresponds to $\tau_e/\tau$. The
growth is characterized by the kinetics of island formation, the value
of the island concentration at saturation $N_{sat}$ (i.e. the maximum
value reached before island-island coalescence becomes important) and
the corresponding values of the thickness $e_{sat}$ and condensation
coefficient $C_{sat}$, useful when evaporation is important (the
condensation coefficient is the ratio of matter actually present on the
substrate over the the total number of particles sent on the surface
(also called the thickness $e=Ft$), see Table I). 

I also give the island size distributions corresponding to each growth
hypothesis. These have proven useful as a tool for experimentalists to
distinguish between different growth mechanisms
\cite{ss,stroscio,evans}. By {\it size} of an island, I mean the
surface it occupies on the substrate. For "two dimensional" islands
(i.e. formed by pure juxtaposition), this is the same as the island
mass, i.e. its number of monomers. For "three dimensional" islands
(formed by total coalescence), their projected surface is the easiest
quantity to measure by microscopy.  It should be noted that for three
dimensional islands, their projected surface for a given mass depends
on their shape, which is assumed here to be pyramidal (close to a
half-sphere). It has been
shown \cite{bartelt_vac,model} that by normalizing the size histograms, one obtains a "universal" size
distribution independent of the coverage, the flux or the
substrate temperature for a large
range of their values.

\subsection{Pure juxtaposition : growth of one cluster thick islands}

I first study the formation of the islands in the limiting case of
pure juxtaposition. This is done for several growth hypothesis. The
rate-equations treatment is given in Appendix A.

\subsubsection{Complete condensation}

Let me start with the simplest case where only diffusion takes place
on a perfect substrate (no evaporation).  Fig \ \ref{dent}a shows the
evolution of the monomer (i.e.  isolated clusters) and island
densities as a function of deposition time.

We see that the monomer density rapidly grows, leading to a rapid
increase of island density by monomer-monomer encounter on the
surface. This goes on until the islands occupy a small fraction
of the surface, roughly 0.1\% (Fig. \ref{morpho}a). Then, islands 
capture efficiently the
monomers, whose density decreases. As a consequence, it becomes less
probable to create more islands, and their number
increases more slowly. When the coverage reaches a value close to
15\% (Fig. \ref{morpho}b), coalescence will start to decrease the 
number of islands. The
maximum number of islands at saturation $N_{sat}$ is thus reached for
coverages around 15\%.  Concerning the dependence of $N_{sat}$ as a
function of the model parameters, it has been
 shown that the maximum number of islands per unit area formed on the
surface scales as $N_{sat} \simeq (F/D)^{1/3}$
\cite{venables73,stoyanov}. Recent simulations \cite{bales,ss} and
theoretical analysis \cite{villain} have shown 
that the precise relation is $N_{sat} = 0.53 (F \tau)^{0.36}$ for the 
{\it ramified} islands produced by pure juxtaposition (Fig. \ \ref{nmax2d}).

It should be noted that if cluster diffusion is vanishingly small, the above
relation does not hold : instead, film growth proceeds as in the percolation
model \cite{perco_model}, by random paving of the substrate. An experimental
example of such a situation has been given in Ref. \cite{percosb}.

\subsubsection{Evaporation}

What happens when evaporation is also included ? Fig \ \ref{dent}b shows
that now the monomer density becomes roughly a {\it constant}, since
it is now mainly determined by the balancing of deposition and
evaporation. As expected, the constant concentration equals $F
\tau_e$ (solid line). Then the number of islands
increases linearly with time (the island creation rate is roughly
proportional to the square monomer concentration, see Appendix A). One
can also notice that
only a small fraction (1/100) of the monomers do effectively remain on
the substrate, as shown by the low condensation coefficient value at early
times.  This can be understood by noting that the islands grow by
capturing only the monomers that are deposited within their "capture
zone" (comprised between two circles of radius $R$ and $R+X_S$).  The
other monomers evaporate before reaching the islands.  When the islands 
occupy a significant
fraction of the surface, they capture rapidly the monomers. This has
two effects : the monomer density starts to decrease, and the condensation
coefficient starts to increase. Shortly after, the island density
saturates and starts to decrease because of island-island coalescence.
Fig. \ \ref{nmax2d} shows the evolution of the maximum island density
in the presence of evaporation. A detailed analysis of the effect of
monomer evaporation on the growth is given in Ref. \cite{evaprb}, where
is also discussed the regime of "direct impingement" which arises
when $X_S \leq 1$ : islands are formed by {\it direct
impingement} of incident clusters as first neighbors of previously adsorbed
clusters, and grow
by direct impingement of clusters on the island boundary. A summary of
the results obtained in the various regimes spanned as the
evaporation time $\tau_e$ decreases is given in Appendix A.

\subsubsection{Defects}

I briefly treat now the influence of a very simple kind of defect : a
perfect trap for the diffusing particles. If a particle enters such a
defect site, it becomes trapped at this site forever.  If such defects are
present on the surface \cite{defect_point} they will affect the growth of
the film only if their number is higher than the number of islands
that would have been created without defects (for the same values of the 
parameters). If this is indeed the case,
monomers will be trapped by the defects at the very beginning of the
growth and the number of islands equates the number of defects,
whatever the diffusivity of the particles. The kinetics of island
formation is dramatically affected by the presence of defects, the
saturation density being reached almost immediately (Fig. \
\ref{cinetns}).

\subsubsection{Island mobility}

The consequences of the mobility of small islands have not received
much attention. One reason is that it is difficult to include island
mobility in rate equations treatments. A different (though related!)
reason is that (atomic) islands are expected to be almost immobile in
most homoepitaxial systems. However, several studies have shown the
following consequences of island mobility for the pure juxtaposition
case and in the absence of evaporation.  First, the saturation island
density is changed \cite{villain,boston,metiu,kuipers,furman} : one 
obtains $N_{sat}= 0.3 (F/D)^{.42}$ (Fig. \ref{nmax2d}) if all islands 
are mobile, with a mobility
inversely proportional to their size \cite{boston}. Second, the
saturation island density is reached for very low coverages (Fig. \
\ref{cinetns} and Ref. \cite{boston}). This can be explained by a
dynamical equilibrium between island formation and coalescence taking
place at low coverages thanks to island diffusion. If only monomers are 
able to move, islands can coalesce (static coalescence) only when the coverage
is high enough (roughly 10-15\%, \cite{villain,evaprb}). Then, the
saturation island density is reached in this case for those
coverages. Instead, when islands can move, the so called dynamical
coalescence starts from the beginning of the growth and the balance is
established at very low coverages \cite{boston}. Third, the island
size distribution is sharpened by the mobility of the islands
\cite{mrs,japan,kuipers,furman}.  To my knowledge, there is no 
prediction concerning the growth of films with evaporation when islands 
are mobile.

\subsubsection{Island size distributions}
 
Fig \ \ref{distrib2d} shows the evolution of the {\it rescaled}
\cite{model,stroscio} island size distributions as a function of the
evaporation time for islands formed by juxtaposition \cite{evaprb}.
Size distributions are normalized by the mean island
size in the following way : one defines $p(s/s_m) = n_s / N_t$ as the
probability that a randomly chosen island has a surface $s$ when the
average surface per island is $s_m = \theta/N_t$, where $n_s$ stands
for the number of islands of surface $s$, $N_t$ is the total number of
islands and $\theta$ for the coverage of the surface.  
It is clear that the distributions are significantly affected by the
evaporation, smaller islands becoming more numerous when evaporation
increases. This trend can be qualitatively understood by noting that
new islands are created continuously when evaporation is present,
while nucleation rapidly becomes negligible in the complete
condensation regime. The reason is that islands are created
(spatially) homogeneously in the last case, because the positions of
the islands are correlated (through monomer diffusion), leaving
virtually no room for further nucleation once a small portion of the
surface is covered ($\theta \sim 0.05$). In the limit of strong
evaporation, islands are nucleated randomly on the surface, the
fluctuations leaving large regions of the surface uncovered. These
large regions can host new islands even for relatively large
coverages, which explains that there is a large proportion of small
($s < s_m$) islands in this regime.

\vspace{2cm}

\subsection{Total coalescence : growth of three dimensional islands}
\label{sec_3d}

If clusters coalesce when touching, the results are slightly different
from those given in the preceding section, mainly because
 the islands occupy a smaller portion of the substrate at a
given thickness. Therefore, in the case of complete condensation
for example, saturation arises at a higher thickness (Fig. \ \ref{cinetns})
even if the {\it coverage} is approximately the same (matter is "wasted"
in the dimension perpendicular to the substrate). However, the main
qualitative characteristics of the growth correspond to those
detailed in the preceding section.
Fig. \ \ref{nmax3d} shows the evolution of the
maximum island density in that case, where the three-dimensional
islands are assumed to be roughly half spheres (actually, pyramids
were used in these simulations which were originally intended for
atomic deposition \cite{henry}). The analytical results obtained from
a rate-equations treatment are given in Appendix B. If the islands 
are more spherical
(i.e. the contact angle is higher), a simple way to adapt these results
on the kinetic evolution of island concentration (Fig. \ \ref{cinetns}) is
to multiply the thickness by the appropriate form factor, 2 for a
sphere for example. Indeed, if islands are spherical, the same
coverage is obtained for a thickness double than that obtained for
the case of half-spheres (there are two identical half spheres). This is a
slight approximation since one has to assume that the capture cross
section (which governs the growth) is identical for the two shapes :
this is not exactly true \cite{evaprb} but is a very good
approximation.

Fig \ \ref{distrib3d} shows the evolution of the {\it rescaled} island
size distributions for three dimensional islands (pyramids) in
presence of evaporation. I recall that size means here the projected
surface of the island, a quantity which can be measured easily by
electron microscopy. We note the same trends as for the pure
juxtaposition case.

Fig \ \ref{distribdef} shows the evolution of the {\it rescaled}
island size distributions for pyramidal islands nucleating {\it on
defects}. Two main differences can be noted. First, the histograms are
significantly narrower than in the preceding case, as had already been
noted in experimental studies \cite{harsdorff}.  This can be
understood by noting that all islands are nucleated at almost the same
time (at the very beginning of growth).  The second point is that the
size distributions are sensitive to the actual coverage of the
substrate, in contrast with previous cases.  In other words, there is
no perfect rescaling of the data obtained at different coverages, even
if rescaling for different fluxes or diffusion times has been checked.

\subsection{Other growth situations}

I briefly address in this paragraph other processes which have
 not been analyzed here. A possible (but difficult to study) process is a long
range interaction between particles (electrostatical or through the
substrate). There is some experimental evidence
of this kind of interaction for the system Au/KCl(100) \cite{zanghi} but 
to my knowledge, it has never been incorporated in growth models. 
Chemical impurities adsorbed on the substrate can change the growth in
conventional vacuum, and these effects are extremely difficult to
understand and control \cite{defects}. Of course, many other possible
processes have not been addressed in this review, such as
the influence of strain, of extended defects as steps or 
vacancy islands \ldots

\section{HOW TO ANALYZE EXPERIMENTAL DATA}
\label{sec_interpret}

Figures \ \ref{nmax2d}, \ref{nmax3d} constitute in some sense
"abacuses" from which one can determine the value of the microscopic
parameters (diffusion, evaporation) if the saturation island density
is known. The problem is : does the measured island density correspond to the
defect concentration of the surface or to homogeneous nucleation? If
the latter is true : which curve should be used to interpret the data?
In other words, is evaporation present in the experiments and what
is the magnitude of $\tau_e$? I will now give some tricks to first find out
which processes are relevant and then how they can be quantified. 

Let's concentrate first on the presence of defects. One possibility is 
to look at the evolution of $N_{sat}$ with the flux. As already explained, 
if this leaves unaffected the saturation density, nucleation is occurring on
defects. A similar test can be performed by changing the substrate temperature,
but there is the nagging possibility that this changes the
defect concentration on the surface. It is also possible to study
the kinetics of island nucleation, i.e. look at the island
concentration as a function of thickness or coverage. The presence of
defects can be detected by the fact that the maximum island density is
reached at very low coverages (typically less than 1\%, see Fig. \
\ref{cinetns}) and/or by the fact that the nucleation rate (i.e. the
derivative of the island density) scales as the flux and not as the square
flux : see Section 3 of Ref. \cite{stoyanov} for more details.  One should be 
careful however to check that all the
islands, even those containing a few particles, are visible in the
microscope images. This is a delicate point for atomic deposition
\cite{visible} but should be less restrictive for clusters since each
cluster has already a diameter typically larger than a nanometer. Of course
all this discussion assumes that the defects are of the "ideal" kind
studied here, i.e. perfect traps. If atoms can escape from the defects
after some time, the situation is changed but I am unaware of studies
on this question.

The question of evaporation is more delicate. First, one should check 
whether particle reevaporation is important. In principle, this can
be done by measuring the
condensation coefficient, i.e.  the amount of matter present on the
surface as a function of the amount of matter brought by the beam. 
If possible, this
measure leaves no ambiguity. Otherwise, the kinetics of island
creation is helpful. If the saturation is reached at low thicknesses
($e_{sat} \leq .5 \ ML$), this means that evaporation is not
important. Another way of detecting particle evaporation is by studying the
evolution of the saturation island density with the
flux : in the case of 2D growth (Fig. \ \ref{nmax2d}), 
the exponent is 0.36 when
evaporation is negligible but roughly 0.66 when evaporation
significantly affects the growth \cite{evaprb}. There are similar
differences for 3d islands : the exponent changes from 0.29
to 0.66 \cite{pyr} (Fig. \ \ref{nmax3d}).  Suppose now that
one finds that evaporation is indeed important :
before being able to use Fig. \ \ref{nmax2d} or Fig. \ \ref{nmax3d},
one has to know the precise value of $\tau_e$. One way to find out 
is to make a precise fit of the kinetic evolution of the island density or
the condensation coefficient (see Section \ref{sec_sb36} for an example). 
In next paragraph, I show how to 
find $\tau_e$ if one knows only the {\it saturation} values of the
island density and the thickness.

\vspace{1cm}

As a summary, here is a possible experimental strategy to analyze
the growth. First, get a series of micrographs of submonolayer films
as a function of the thickness. The distinction between the 
pure juxtaposition and total coalescence cases can be
easily made by comparing the size of the 
supported islands to the (supposedly known) size of the incident 
clusters. Also, if the islands
are spherical, this means that coalescence has taken place, if they
are ramified that clusters only juxtapose upon contact. Of course, all
the intermediate cases are possible (see the case of gold clusters
below). One can calculate the ratio of deposited thickness over the 
coverage : if this ratio is close to 1, islands are flat (i.e. one
cluster thick), otherwise 
three dimensional (unless there is evaporation). 

From these micrographs, it is possible to measure the island density
as a function of the thickness. Fig. \ \ref{cinetns} should be now
helpful to distinguish between the different growth mechanisms. For
example, if the saturation island density is obtained for large
thicknesses (typically more than 1ML), then evaporation is certainly
relevant and trying to measure the condensation coefficient is important
to confirm this point.  It is clear from Figs. \ \ref{nmax2d}, \
\ref{nmax3d} that the knowledge of $N_{sat}$ alone cannot determine
$\tau_e$ since many values of $\tau$ and $\tau_e$ can lead to the same
$N_{sat}$. In the 2D case, the values of the microscopic parameters
can be obtained by noting that the higher the evaporation rate, the
higher the amount of matter "wasted" for film growth (i.e. re-evaporated).
One therefore expects that the smaller $\tau_e$, the higher $e_{sat}$, which
is confirmed by Fig. \ \ref{coll2d}a. Therefore, from the (known) value of
 $e_{sat}$, one can determine the value of the evaporation parameter 
$\eta = F \tau X_S^6$ (Fig. \ \ref{coll2d}a).
Once $\eta$ is known, $X_S$ is determined from Fig. \ \ref{coll2d}b since
$N_{sat}$ is known. $F\tau$ can afterwards be determined (from $X_S$ and 
$\eta$). This is only valid for $X_S \gg 1$ \protect\cite{evaprb}, a 
condition always fulfilled in experiments.

The 3d case is more difficult
since the same strategy (measuring $N_{sat}$ and $e_{sat}$) fails. 
The reason is that in the limit of high evaporation, $e_{sat}$ goes as
$e_{sat} \sim {N_{sat}}^{-1/2}$, thus bringing no independent 
information on the parameters \cite{pyr}. The same is true for the 
condensation coefficient at saturation $C_{sat}$, which is a {\it constant}, 
i.e. independent of the value of $\tau_e$ or the normalized flux (see
Fig. \ \ref{coll3d}b). This 
counterintuitive result (one would think that the higher the evaporation 
rate, the smaller the condensation coefficient at saturation) can be
understood by noting that {\it in this limit}, islands only grow by direct
impingement of particles within them \cite{pyr} and therefore $X_S$ (or
$\tau_e$) has no effect on the growth. Fortunately, in 
many experimental situations, the limit of high evaporation is not reached 
and one "benefits" from (mathematical) crossover regimes where these 
quantities do depend on the precise values of $\tau_e$.  Figs. \ \ref{coll3d} 
give the evolutions of $C_{sat}$ and  $e_{sat}$ as a function of
$N_{sat}$ for different values of $\tau_e$ and F. Then, knowing $e_{sat}$
and $N_{sat}$ leads to an estimation of $\tau_e$ from Fig. \ \ref{coll3d}a
which can be confirmed with Fig. \ \ref{coll3d}b provided $C_{sat}$ is
known. 

To conclude, let me note that a saturation thickness much smaller 
than 1ML can also be attributed to island mobility. This is a subtle
 process and it is difficult to obtain any 
information on its importance. We note that
interpreting data as not affected by island diffusion when it is
actually present leads to errors on diffusion coefficients of one
order of magnitude or more depending on the value of $F\tau$ (see
Fig. \ \ref{nmax2d}). Finally, one should be careful in interpreting
the $N_t$ vs. thickness curves since most observations are not made in
real time (as in the computer simulations) and there can be
post-deposition evolutions (see for example Ref. \cite{bruneprl} for
such complications in the case of atomic deposition).

\section{EXPERIMENTAL RESULTS}
\label{exps}

I review in this section the experimental results obtained these last
years for low-energy cluster deposition, mainly in the submonolayer
regime. The scope is double : first,
I want to give some examples on how to analyze experiments (as indicated in
Section \ref{sec_interpret}) and second, I will show that from a comparison
of experiments and models one can deduce important physical quantities
characterizing the interaction of a cluster with a surface (cluster
diffusivity) and with another cluster (coalescence). The following 
can be read with profit by those interested only in atomic
deposition as examples of interpretation since these elementary
processes are relevant for some cases of atomic deposition. One 
should be careful that some mechanisms which are specific to atomic 
deposition (transient mobility, funnelling, \ldots) are not discussed
here (see \cite{thiel}). Also, growth without cluster diffusion has to 
be interpreted in the framework of the percolation model as indicated
above \cite{percosb}.

Before analyzing experimental data, it is important to know how to
make the connection between the units used in the programs and the
experimental ones (see also Table I). In the program, the unit length 
is the diameter of a cluster. In the experiments, it is therefore 
convenient to use as a
surface unit the {\it site}, which is the projected surface of a
cluster $\pi d^2/4$ where $d$ is the mean {\it incident} cluster
diameter.  The flux is then expressed as the number of clusters
reaching the surface per second per site (which is the same as ML/s) 
and the island density is given per
site. The thickness is usually computed in cluster monolayers (ML),
obtained by multiplying the flux by the deposition time. The coverage -
the ratio of the area covered by the supported islands over the
total area - has to be measured on the micrographs.

\subsection{A simple case : $Sb_{2300}$ clusters on graphite HOPG}
\label{sb2300}

I start with the case of antimony clusters containing 2300 ($\pm 600$) atoms 
deposited on graphite HOPG since here the growth has been thoroughly
investigated \cite{ss}. I first briefly present the experimental
procedure and then the results and their interpretation in terms of 
elementary processes.

\subsubsection{Experimental procedure}

As suggested in the preceding Section, various samples are prepared 
for several film thicknesses, incident fluxes and the substrate temperatures. 
For films grown
on Highly Oriented Pyrolitic Graphite (HOPG), before deposition at
room temperature, freshly cleaved graphite samples are annealed at
500$^\circ$C during 5 hours in the deposition chamber (where the
pressure is $\simeq 10^{-7}$ Torr) in order to clean the surface. The 
main advantage of HOPG graphite
conveniently annealed is that its surfaces consist mainly of
defect-free large terraces ($\simeq 1\mu m$) between steps. It is also
relatively easy to observe these surfaces by electron or tunneling
microscopy \cite{ss}.  Therefore, deposition on graphite HOPG is a
good choice to illustrate the interplay between the different
elementary processes which combine to lead to the growth. After
transfer in air, the films are observed by Transmission Electron
Microscopy (TEM) (with JEOL 200CX or TOP CON electron microscopes
operating at 100 kV in order to improve the contrast of the
micrographs).

\subsubsection{Results}

Fig. \ \ref{sb}a shows a general view of the morphology of the
antimony submonolayer film for $e=0.14$ ML and $T_s=353K$.  A detailed
analysis \cite{ss} of this kind of micrographs shows that the ramified
islands are formed by the juxtaposition of particles which have the
same size distribution as the free clusters of the beam. From this, we
can infer two important results.  First, clusters do not fragment upon
landing on the substrate as indicated in the introduction. Second,
antimony clusters remain juxtaposed upon contact and do not coalesce
to form larger particles (option (a) of Fig. \ \ref{interaction}).

From a qualitative point of view, Fig. \ \ref{sb}a also shows that the
clusters are able to {\it move} on the surface.  Indeed, since the
free clusters are deposited at random positions on the substrate, it
is clear that, in order to explain the aggregation of the clusters in
those ramified islands, one has to admit that the clusters move on the
surface. How can this motion be quantified? Can we admit that diffusion
and pure juxtaposition are the only important physical phenomena at
work here?

Fig. \ \ref{sbss}a shows the evolution of the island density as a
function of the deposited thickness. We see that the saturation island
density $N_{sat}$ is reached for $e \simeq 0.15 ML$. This indicates
that evaporation or island diffusion are not important in this
case. Therefore, we guess that the growth should be described by a
simple combination of deposition, diffusion of the incident clusters
and juxtaposition. This has been confirmed in several ways. I only
give three different confirmations, directing the reader to
Ref. \cite{ss} for further details. First, a comparison of the
experimental morphology and that predicted by models including only
deposition, diffusion and pure juxtaposition shows a very good
agreement (Fig. \ \ref{sb}b). Second, Fig. \ \ref{sbss}b shows that
the saturation island density accurately follows the prediction of the
model when the flux is varied. I recall that if the islands were
nucleated on defects of the surface, the density would not be
significantly affected by the flux.  

Having carefully checked that the experiments are well described by
the simple DDA model, I can confidently use Fig. \ \ref{nmax2d} to
quantify the diffusion of the clusters. As detailed in Ref. \cite{ss},
one first measures the saturation island density for different substrate
temperatures. The normalized fluxes ($F\tau$) are obtained from Fig. \ \ref{nmax2d}. Knowing the experimental fluxes, one can derive the
diffusion times and coefficients.  The result is a surprisingly high
mobility of $Sb_{2300}$ on graphite, with diffusion coefficients of
the same order of magnitude as the atomic ones, i.e. $10^{-8} cm^{2}
s^{-1}$ (Fig. \ \ref{sbss}c).

The magnitude of the diffusion coefficient is so high that we wondered
whether there was any problem in the interpretation of the data, in despite
of the very good agreement between experiments and growth models described
above. For example, one could think of a linear diffusion of the incoming
clusters, induced by the incident kinetic energy of the cluster in the beam 
(the cluster could "slide" on the graphite surface). This seems unrealistic
for two reasons : first, in order to explain the low island density obtained 
in the experiments (see above), it should be assumed that the cluster, 
which has a low kinetic energy (less than 10 eV), can travel at least several
thousand nanometers before being stopped by friction with the substrate. This
would imply that the diffusion is just barely influenced by the substrate, 
which only slows down the cluster. In this case, it is difficult to explain 
the large changes observed in the island density when the substrate 
temperature varies. Second, we have deposited antimony clusters on a 
graphite substrate tilted to $30 ^{o}$ from its usual position (i.e. 
perpendicular to the beam axis). Then, a linear diffusion of the 
antimony clusters arising from their incident kinetic energy would lead
to anisotropic islands (they would grow differently in the direction of 
tilt and its perpendicular). Experiments \cite{ss} show that there is no
difference between usual and tilted deposits. Therefore we can confidently
believe that $Sb_{2300}$ clusters perform a very rapid Brownian motion 
on graphite surfaces.  A similar study has been carried out for $Sb_{250}$ 
on graphite, showing the same order of magnitude for the mobility of the 
clusters \cite{ss}. The microscopic mechanisms that could explain such a
motion will be presented in section \ref{sec_atomic}.

\subsection{Other experiments}
\label{otherexp}

In this subsection, I try to analyze data obtained in previous
studies \cite{palmerdepo,thesefco}. I provide {\it possible} 
(i.e. not in contradiction
with any of the data) explanations, with the respective values of the
microscopic processes. I stress that the scope here is not to make
precise fits of the data, but rather to identify the
microscopic processes at work and obtain good guesses about their
respective values.

\subsubsection{Slightly accelerated $Ag_{160}$ clusters on HOPG}
\label{ag}

Palmer's group \cite{palmerdepo} has investigated the growth of films
by $Ag_{160}$ cluster deposition. Fig. \ref{palmerag} shows
the ramified morphology of a submonolayer deposit. Although no
precise fit is possible given the limited experimental
data, the island density and size shows that $Ag_{160}$ clusters 
are mobile on HOPG.

\subsubsection{$Sb_{36}$ on a-C}
\label{sec_sb36}

Small antimony clusters are able to move on amorphous
carbon, as demonstrated by Figs. \ \ref{sb36}, and by the fact that
the films are dramatically affected by changing the incident flux
\cite{thesefco}.

Fig. \ \ref{sb36}a shows that these small clusters gather in large
islands and coalesce upon contact. The island density is shown in
Fig. \ \ref{sb36}b. The maximum is reached for a very high thickness
($e \simeq 1.8$ML), which can only be explained by supposing that
there is significant reevaporation of $Sb_{36}$ clusters from the
surface. Evaporation of small antimony clusters ($Sb_n$ with n $\leq$
100) from a-C substrates has also been suggested by other authors
\cite{brechignac}. A fit, using $\tau_e = 20$ deduced from Fig. \
\ref{coll3d}a gives, with $F\tau=10^{-5}$ for $F=6 \ 10^{-3}$
clusters \ $site^{-1} s^{-1}$, leading to $\tau \sim 2 \ 10^{-3}s, D= 2
\ 10^{-12} cm^2/s, \tau_e = .04 s$ and $X_S \sim 6$ nm before
evaporation, and a condensation coefficient of .2 when the maximum island
density is reached. However, some authors have argued \cite{fuchs}
that the condensation coefficient is not so low. It is interesting to try
a different fit of the data -- in better agreement with this
indication -- to give an idea of the uncertainties of the fits. For
this, I assume that the deposited islands are spherical (solid line of
Fig. \ \ref{sb36}b) by the procedure described in Section \
\ref{simu}.  Here I have taken $F\tau=3 \ 10^{-6} for f=6 \ 10^{-3}$
clusters \ $site^{-1} s^{-1}$, leading to $\tau \sim 5 \ 10^{-4}s,
\tau_e = .04 s$ corresponding to $D= 8 \ 10^{-12} cm^2/s$, and $X_S
\sim 11$ nm before evaporation, and a condensation coefficient of .5 when
the maximum island density is reached. Note that the condensation
coefficient is, as expected, higher than in the previous fit and that
the agreement with the experimental island densities for the lowest
thicknesses is better. Comparing the two fits, it can be seen that the
difference in the diffusion coefficient is a factor of 4, and a factor
2 in the $X_S$.  This means that the orders of magnitude of the values
for the microscopic mechanisms can be trusted despite lack of
comprehensive experimental investigation.

Similar studies  \cite{moi_unpub} have allowed to obtain the diffusion and 
evaporation characteristic times for other clusters deposited on amorphous 
carbon. For $Bi_{90}$, one finds $D \sim 3 \ 10^{-13} cm^{2} s^{-1}$ and 
$X_S \sim 8$ nm and for $In_{100}$ : $D \simeq 4 \ 10^{-15} cm^2 s^{-1}$ 
and strong coalescence (the incident clusters are liquid).

\subsubsection{$Au_{250}$ on graphite}
\label{fhach}

Fig. \ \ref{au} shows the morphology of a gold submonolayer film
obtained by deposition of $Au_{250}$ ($\pm 100$) clusters prepared by a laser
source on graphite in a UHV chamber for different substrate temperatures.

The structures are strikingly similar to those obtained in the
$Sb_{2300}$ case : large, ramified islands.  We can conclude that
$Au_{250}$ clusters do move on graphite, and that they do not
completely coalesce. A more careful examination of the island
morphology indicates that the size of the branches is
not the same as the size of the incident clusters, as was the case for
$Sb_{2300}$.  Here the branches are larger, meaning that there is a
partial coalescence, limited by the {\it kinetics} of the growth. This
is a very interesting experimental test for coalescence models that
are presented later. I first try to estimate the diffusion coefficient
of the gold clusters. We have to be careful here because the incident
flux is {\it chopped} with the laser frequency, roughly 10Hz.  The
active portion of the period (i.e. when the flux is "on") is $\simeq
100\mu s$.

An analysis of the growth in presence of a chopped flux has been
reported elsewhere \cite{hache,ncombe}. Fig. \ \ref{denhache}a shows
the values of $N_{sat}$ as a function of the diffusion time $\tau$ in
these experimental conditions ($F_i$=6 ML/s) for two hypothesis : only
the monomers move or islands up to the pentamer move too (island
mobility is supposed to be inversely proportional to its mass). Note
that there is a range of diffusion times (up to two orders of
magnitude) which lead to the same island saturation value, a strange
situation in homogeneous nucleation : see Refs. \cite{hache,ncombe}
for details.

Given the experimental island densities, the diffusion coefficients in
both hypothesis are shown in Fig. \ \ref{denhache}b. The values of the
diffusion coefficient seem too high, especially in the case of
exclusive monomer diffusion, but there is no experimental evidence of
island mobility for the moment. I note however that since the incident
clusters do significantly coalesce, it is not unreasonable to assume
that the smallest islands (which are spherical as the incident
clusters) can move too. We are presently  carrying additional tests 
(on cluster reevaporation or non brownian cluster diffusion) to confirm
the observation of such high diffusion coefficients.

\subsubsection{$Au_{250}$ on $NaCl$}

Given the surprising high mobility of $Au_{250}$ ($\pm 100$) on HOPG, 
it was worth
testing gold cluster mobility on other substrates. I present here recent
results obtained by depositing $Au_{250}$ clusters on $NaCl$
\cite{aunacl}. The high island density (Fig. \ \ref{aunaclmorph}a)
shows that gold clusters are not very mobile on this substrate, with
an upper limit on the diffusion coefficient of $D \simeq
10^{-15}cm^2/s$.  This is in agreement with the low mobilities
observed by other authors \cite{massonexp} in the seventies. 
The diffraction pattern (Fig. \ \ref{aunaclmorph}b) is similar to that 
obtained in Figs. 15 c and d
of Ref. \cite{massonexp}. The authors interpreted their results with the
presence of multi-twinned Au particles with two epitaxial orientations :
Au(111)/NaCl(100) and Au(100)/NaCl(100). This is reasonable taking into
account the interatomic distances for these orientations : 
$d_{Au_Au(111)}$=0.289nm, \ $d_{Na_Cl(100)}$=0.282nm and  
$d_{Au_Au(100)}$=0.408nm \ $1/2 d_{Na_Cl(100)}$=0.398nm (along the
face diagonal). These preliminary
results suggest that epitaxy may prevent clusters from moving rapidly
on a surface, a result which has also be observed by other groups
(see next section). They also show that, at least in this
case, forming the clusters on the surface by atomic aggregation or
depositing preformed clusters does not change the orientation nor the
diffusion of the clusters on the surface. Work is in progress to determine
the precise atomic structure of the clusters, their orientation on the
substrate and their diffusion at higher temperatures \cite{aunacl}.

\section{TOWARDS A PICTURE OF CLUSTER DIFFUSION AND COALESCENCE 
AT THE ATOMIC SCALE}
\label{sec_atomic}

In the preceding sections I have tried to analyze the growth with the
help of two main ingredients : diffusion of the clusters on the
surface and their interaction. I have
taken the diffusion as just one number quantifying the cluster motion, 
without worrying about the microscopic mechanisms which could explain it. 
For {\it atomic} diffusion, these mechanisms have been extensively
studied \cite{lagally,thiel,gomer} and are relatively well-know. In the
(simplest) case of compact (111) flat surfaces, diffusion occurs by site to 
site jumps over bridge sites (the transition state). Therefore, diffusion 
is an activated process and
plotting the diffusion constant vs. the temperature yields the height
of the barrier, which gives information about the microscopics of diffusion.
This kind of simple interpretation is not valid for {\it cluster} diffusion.
It is always possible to infer an "activation" energy from an Arrhenius plot
(see Fig. \ \ref{sbss}c) but the meaning of this energy is not clear since
the precise microscopic diffusion mechanism is unknown. 

Similarly, cluster-cluster coalescence (Fig. \ \ref{interaction}) has been 
supposed to be total or null (i.e. pure
juxtaposition) but without considering the kinetics nor the
intermediate cases which can arise (see the experimental results for
gold on graphite for example).

In this section, I describe some preliminary results which can shed some 
light on the microscopic mechanisms leading to cluster diffusion or coalescence.

\subsection{Diffusion of the clusters}

Before turning to the possible microscopic
mechanisms, one must investigate whether cluster diffusion is indeed such a
general phenomenon. Let me review now the available experimental data 
concerning the diffusion of {\it three-dimensional} (3D) clusters. I
already presented in the previous section several examples of high cluster
mobilities over surfaces. In the case of $Sb_{2300}$ on graphite, mobilities 
as high as $D = 10^{-8} cm^{2} s^{-1}$ are obtained at room temperature, and
similar values can be inferred for Ag cluster deposition 
\cite{palmer,palmerdepo}. 
On a-C substrates, diffusion is not that rapid, but has to be taken into 
account to understand the growth. More than twenty years ago, the 
Marseille group \cite{zanghi,massonexp,metois,kern}
carefully studied the mobility of nanometer-size gold crystallites on
ionic substrates (MgO, KCl, NaCl). By three different methods, they proved
that these 3D clusters - grown by atomic deposition at room temperature - 
are significantly mobile at moderately high 
temperatures ($T \sim 350 K$). The three different methods were : direct 
observation under the electron microscope beam \cite{metois}, comparison of
abrupt concentration profiles \cite{massonexp} or the radial distribution 
functions \cite{zanghi} before and after annealing. All these results are 
carefully reviewed in Ref. \cite{kern}. I will focus here on the last
method \cite{zanghi}. Fig. \ref{radial} shows the radial distribution 
functions of the gold clusters obtained just after deposition (the flat curve)
and after annealing (the oscillating curve) a similar deposit for a few 
minutes at 350K (Fig. 4 of Ref. \cite{zanghi}). The flat curve is a standard 
as-grown radial
distribution function (see for example Ref. \cite{model}). The other curve is
significantly different from the first, although the cluster size distribution
remains identical (Fig. \ref{radial}). This shows that gold clusters move
as an entity on KCl(100) at 350K, since the conservation of the size
distribution rules out atomic exchange between islands (the (EC) mechanism
presented below (Section \ref{sec_diff2d}). From the shape of the radial
distribution function some features of the cluster-cluster interaction
could be derived, mainly that it is a {\it repulsive} interaction. The
detailed interaction mechanisms are not clear \cite{zanghi,kern}. A 
different study \cite{massonexp} showed that the clusters were mobile only 
for a limited amount of time (several minutes), and
then stopped.  It turns out that clusters stop as soon as they reach
epitaxial orientation on the substrate. Indeed, the
gold(111) planes can orient on the KCl(100) surface, reaching a stable,
minimum energy configuration (for more details on the epitaxial orientations
of gold clusters on $NaCl$, see Refs. \cite{matthews,kuo}). 
Therefore, 3D cluster diffusion might be quite a common phenomenon, at 
least when there is no epitaxy between the clusters and the substrate. 

What are the possible microscopic mechanisms? Unfortunately for the
field of cluster {\it deposition}, recent theoretical
and experimental work has focused mainly on one atom thick, two-dimensional 
islands, whose diffusion mechanisms might be different from those of
3D islands. The focus on 2D islands is due to the technological impetus
provided by applications of atomic deposition - notably MBE for which one wants
to achieve flat layers. Let's briefly review the current state of the understanding of 2D island diffusion to see what inspiration we can draw for
3D cluster diffusion. 

\subsubsection{2D island diffusion mechanisms}
\label{sec_diff2d}

There are two main types of mechanisms proposed to account for 2D island
diffusion : single adatom motion and collective (simultaneous) atom motion.
It should be noted that small islands (less than $\sim $ 15 atoms) are likely
to move by specific mechanisms, depending on the details of the island
geometry and atomic energy barriers \cite{kellprl,kellogg,smallclu}. 
Therefore I concentrate here on larger 2D islands.

\vspace{1cm}

\paragraph{Individual mechanisms}

\vspace{.4cm}

The most common mechanism invoked to account for 2D island diffusion
has been that of individual atomic motion. By individual I mean that 
the movement of the whole island can be
decomposed in the motion of {\it uncorrelated} single atom moves. There
are two main examples of such a diffusion : evaporation-condensation (EC)
and periphery-diffusion (PD). Theoretical investigations on these individual
mechanisms have generated much interest since it was conjectured that this
diffusion constant $D_{ind}$ is proportional to the island number of atoms
(island mass) to some power which depends on the precise mechanism (EC or PD)
causing island diffusion but not on temperature or the chemical nature of 
the system. If true, this
conjecture would prove very useful, for it would allow to determine
experimentally the mechanism causing island migration by measuring the
exponent and some details of the atom diffusion energetics by measuring how 
$D_{ind}$ depends on temperature. Unfortunately, recent studies have
shown that this prediction is too simplistic, as I show now for the
two different mechanisms. 

\vspace{.4cm}

{\it (i) Periphery diffusion}

\vspace{.5cm}

Fig. \ref{PD} shows the elementary mechanism leading to island diffusion
via atomic motion on the edge of the island (label PD).
Assuming \cite{metiudiff} that 

- each atomic jump displaces the center of mass of the island by a distance
of order 1/N (where N is the number of atoms of the island),

- each edge atom (density $n_s$) jumps with a rate 
$k \sim exp(-E_{ed}/k_B T)$ where $E_{ed}$ is the activation energy for 
jumping from site to site along the border and $k_B$ is Boltzmann constant

One obtains \cite{metiudiff} :

\begin{equation}
D_{ind} \sim k n_s 1/N^2 \sim exp(-E_{ed}/k_B T) N^{-3/2}
\label{diffpd}
\end{equation}

\noindent
if one postulates that $n_s$, the mean concentration of edge atoms 
is proportional
to the perimeter of the island (i.e. to $N^{1/2}$). This equation allows
in principle to determine the edge activation energy
by measuring the temperature dependence of $D_{ind}$. 

However, recent experiments \cite{pai} and Kinetic Monte-Carlo simulations
\cite{voter,metiudiff,diffpdkmc,sholl} have suggested that 
Eq. \ref{diffpd} is wrong. First, the size exponent is not universal but 
depends on the precise energy barriers for atomic motion (and therefore on
the chemical nature of the material) and, second, the measured activation 
energy does not correspond to
the atomic edge diffusion energy. The point is that the limiting mechanism
for island diffusion is {\it corner} breaking,
for islands would not move over long distances simply by edge diffusion of
the outer atoms \cite{metiudiff}. Further studies are needed to fully
understand and quantify the PD mechanism.

\vspace{.4cm}

{\it (ii) Evaporation-Condensation}

\vspace{.4cm}

An alternative route to diffusion is by exchange of atoms between the
island and a 2D atomic gas. This is the usual mechanism
leading to Ostwald ripening \cite{ostwald}. Atoms can randomly evaporate
from the island and atoms belonging to the 2D gas can condensate on it 
(Fig. \ref{PD}). This leads to fluctuations in the position of the
island center of mass, which are difficult to quantify because of the
possible {\it correlations} in the atomic evaporation and condensation. 
Indeed, an atom which has just evaporated form an island is likely to
condensate on it again, which cannot be accounted by a mean-field theory
of island-gas exchange of atoms \cite{ecth}. The latter leads to 
a diffusion coefficient scaling as the inverse radius of the island 
\cite{ecexp}, while correlations cause a slowing down of diffusion, which 
scales as the inverse {\it square} radius of the island 
\cite{khare,sholl,ecth}.

Experimentally, Wen et al. \cite{ecexp} have observed by STM the movement
of Ag 2D islands on Ag(100) surfaces. They measured a diffusivity almost
independent of the island size, which rules out the PD mechanism and roughly
agrees with their \cite{ecexp} calculation of the size dependence of the EC
mechanism. Since this calculation has been shown to be only approximate, further
theoretical and experimental work is needed to clarify the role of EC in 
2D island diffusion. However, the work by Wen et al.  \cite{ecexp} has
convincingly shown that these islands move significantly and that, for silver,
island diffusion is the main route to the evolution of the island 
size distribution,
contrary to what was usually assumed (Ostwald ripening exclusively due to 
atom exchange between islands, via atom diffusion on the substrate).

\vspace{2cm}

\paragraph{Collective diffusion mechanisms}

\vspace{.5cm}

These individual mechanisms lead in general to relatively slow diffusion of
the islands (of order $10^{-17} cm^2/s$ at room temperature \cite{ecexp}).
For small clusters, different (and faster) mechanisms such as dimer 
shearing, involving the simultaneous displacement of a dimer, have 
been proposed \cite{shi}. More generally,
Hamilton et al. \cite{homohamil} have proposed a different 
mechanism, also involving {\it collective} motions of the atoms, which leads
to {\it fast} island motion. By collective I mean that island 
motion is due to a simultaneous (correlated) motion of (at least) several atoms 
of the island.

Specifically, Hamilton et al. \cite{homohamil} proposed that {\it dislocation} motion could cause rapid diffusion of relatively small (5 to 50 atoms) 
{\it homoepitaxial} islands on fcc(111) surfaces.
Fig. \ref{disloc} shows the basic idea : a row of atoms move simultaneously
from fcc to hcp sites, thus allowing the motion of the dislocation and 
consequently of the island center of mass. Alternative
possibilities suggested by Hamilton et al. for dislocation mediated island
motion are the "kink" mechanism (the same atomic row moves by {\it sequential}
but correlated atomic motion) or the "gliding" mechanism studied below, 
where {\it all} the
atoms of the island move simultaneously. Molecular Dynamics simulations,
together with a simple analytical approach \cite{homohamil} suggest that for
the smallest islands ($N < 20$) the gliding mechanism is favored, for
intermediate sizes ($20 < N < 100$) the dislocation motion has the lowest
activation energy, while for the largest studied islands ($N > 100$) the
preferential mechanism is that of "kink" dislocation motion. It is
interesting to quote at this point recent direct observations of cluster 
motion by field ion microscopy \cite{ehrlich}. Fig. \ref{ir19} shows
successive images of a compact $Ir_{19}$ cluster moving on Ir(111). 
By a careful study, the authors have ruled out the individual atomic
mechanisms discussed above as well as the dislocation mechanism. Instead,
they suggest that gliding of the cluster as a whole
is likely to explain the observed motion \cite{ehrlich}.

Hamilton later studied the case of {\it heteroepitaxial}, strained
islands \cite{hetehamil}.
He has shown that - due to the misfit between the substrate and the island
structures - there can exist islands for which introducing a dislocation
does not cost too much extra energy. These metastable misfit dislocations 
would propagate easily within the islands, leading to "magic" island sizes with 
a very high mobility \cite{hetehamil}.

\vspace{2cm}

\subsubsection{3D island diffusion mechanisms}

For the 3D clusters, the
three microscopic mechanisms presented above are possible in principle.
However, as noted above, the individual atom mechanisms lead to a
diffusivity smaller than the diffusion of $Sb_{2300}$ on graphite 
by several orders of magnitude. These mechanisms have also been
ruled out for the diffusion of gold crystallites on ionic substrates 
\cite{kern}. Several tentative explanations based on the gliding of the 
cluster as a whole over the substrate have been proposed \cite{kern}. 
Reiss \cite{reiss} showed that, for a {\it rigid} crystallite which is 
not in epitaxy on the substrate, the activation energy for rotations might 
be weak, simply
because during a rotation, the energy needed by atoms that have to climb up 
a barrier is partially offset by the atoms going into more stable positions.
Therefore, the barrier for island diffusion is of the same order as that
for an atom, as long as the island does not reach an epitaxial orientation. 
Kern et al. \cite{kern,massonth} allowed for a partial rearrangement of the
interface between the island and the substrate when there is a misfit. The
interface would be composed of periodically disposed zones in registry with
the substrate, surrounded with perturbed ("amorphous") zones, weakly bound
to the substrate. This theory - similar in spirit to
the dislocation theory proposed by Hamilton \cite{homohamil,hetehamil}
for 2D islands - leads to reasonable predictions \cite{kern} 
but is difficult to test quantitatively.

To clarify the microscopic mechanisms of 3D cluster diffusion, I now
present in detail Molecular Dynamics (MD) studies 
carried out recently \cite{mddiff}. 
These simulations aimed at clarifying the generic aspects of the question 
rather than modeling a particular case.
Both the cluster and the substrate are made up of Lennard-Jones atoms \cite{lj},
interacting through potentials of the form :

$ V(r)=4 \epsilon \left (\left(\frac{\sigma}{r} \right)^{12} - 
\left(\frac{\sigma}{r} \right)^{6} \right ) $.  

Empirical potentials of this type, originally
developed for the description of inert gases, are now commonly used to
model generic properties of condensed systems.  Lennard-Jones potentials
include only pair atom-atom interaction and ensure a repulsive interaction
at small atomic distances and an attractive interaction at longer
distances, the distance scale being fixed by $\sigma$ and the energy
scale by $\epsilon$. For a more detailed discussion of the different
interatomic potentials available for MD simulations and their respective
advantages and limitations, see Ref. \cite{mrs_potentials}. The substrate is
modeled by a single layer of atoms on a triangular lattice, attached
to their equilibrium sites by weak harmonic springs that preserve
surface cohesion.  The Lennard-Jones parameters for
cluster atoms, substrate atoms and for the interaction between the
substrate and the cluster atoms are respectively $\left
(\epsilon_{cc},\sigma_{cc} \right)$, $\left (\epsilon_{ss},\sigma_{ss}
\right)$ and $\left (\epsilon_{sc}, \sigma_{sc} \right)$.
$\epsilon_{cc}$ and $\sigma_{cc}$ are used as units of energy and
length.  $\epsilon_{sc}$, $\sigma_{ss}$ and $T$, the temperature of
the substrate, are the control parameters of the simulation.  The last
two parameters are then constructed by following the standard
combination rules : $ \epsilon_{ss} = \sigma_{ss}^{6} $ and $
\sigma_{sc} = \frac {1}{2} \left ( \sigma_{cc} + \sigma_{ss} \right)$. 
Finally, the unit of time is defined as
$\tau = (M \sigma_{cc}^2/\epsilon_{cc})^{1/2}$, where $M$ is the mass
of the atoms which is identical for cluster and substrate atoms. The
simulation uses a standard molecular dynamics technique with
thermostatting of the {\it surface} temperature \cite{A&T}.

In these simulations, the clusters take the spherical cap shape of a 
solid droplet (Fig. \ \ref{mdclu}) partially wetting the substrate. 
The contact angle, which
can be defined following reference \cite{angle}, is roughly
independent of the cluster size (characterized by its number of 
atoms $n$, for $50 < n < 500$. This angle can be tuned by changing 
the cluster-substrate interaction. For large enough $\epsilon_{sc}$, 
total wetting is observed. The results presented below have been obtained 
at a reduced temperature of 0.3 for which the cluster is solid.  This is 
clearly visible in Fig. \ \ref{mdclu}, where
the upper and lower halves of the cluster, colored white and grey at
the beginning of the run, clearly retain their identity after the
cluster {\it center of mass} has moved over 3 lattice
parameters. Hence the cluster motion appears to be controlled by {\it
collective} motions of the cluster as a whole rather than by single
atomic jumps.

The MD simulations have confirmed that one of the most important 
parameters for determining the cluster
diffusion constant is the ratio of the cluster lattice parameter to
the substrate lattice parameter. The results for the diffusion
coefficient are shown in Fig. \ \ref{simudif}a.  When the substrate
and cluster are commensurate ($\sigma_{ss}=\sigma_{cc} \equiv 1$), the cluster
can lock into a low energy epitaxial configuration. A global
translation of the cluster would imply overcoming an energy barrier
scaling as $n^{2/3}$, the contact area between the cluster and the
substrate.  In that case diffusion will be very
slow, unobservable on the time scale of the MD simulations. What is
interesting to note is that even small deviations from
this commensurate case lead to a measurable diffusion on the time
scale of the MD runs.  This can be understood from the fact that the
effective potential in which the center of mass moves is much weaker,
as the cluster atoms, constrained to their lattice sites inside the
rigid solid cluster, are unable to adjust to the substrate potential (see
above, Reiss model \cite{reiss}). The effect is rather spectacular :  
a 10\% change on the lattice parameter induces an increase of  the 
diffusion coefficient by several orders of magnitude.

Finally, I show in Fig. \ \ref{simudif}b the effect of
cluster size on the diffusion constant for different
lattice parameter values.  As the number $n$ of atoms in the cluster
is varied between $n=10$ and $n=500$, the diffusion constant
decreases, roughly following a power law $D\sim n^{\alpha}$.  
This power law exponent $\alpha$ depends
significantly on the mismatch between the cluster and the substrate
lattice parameters.  For high mismatches ($\sigma_{ss}=0.7,0.8$),
$\alpha$ is close to $-0.66$. As the diffusion constant is inversely
proportional to the cluster-substrate friction coefficient, this
result is in agreement with a simple "surface of contact" argument
yielding $D\sim n^{-2/3}$. On the other hand, when the lattice
mismatch is equal to $0.9$, one obtains $\alpha \approx -1.4$,
although the shape of the cluster, characterized by the contact angle,
does not appreciably change.  It is instructive to follow the trajectory
followed by the cluster center of mass (Fig. \ \ref{trajectory}). In 
the runs with a large
mismatch (Fig. \ \ref{trajectory}a), this trajectory is "brownian-like", 
with no apparent influence of the substrate. This is consistent with the 
simple "surface of contact" argument.
Instead, when the mismatch is small (Fig. \ \ref{trajectory}b), the center
of mass of the cluster follows a "hopping-like" trajectory, jumping
from site to site on the honeycomb lattice defined by the
substrate. When $\sigma_{ss}=\sqrt{3}/2$, there seems to be a
transition between the two regimes around $n=200$.

It is interesting to consider the interpretation of cluster motion in terms of
dislocation displacement within the cluster, a mechanism which has been
proposed to explain rapid 2D cluster diffusion \cite{homohamil,hetehamil} 
(see the discussion in Section \ref{sec_diff2d}). 
For this, one can "freeze" the internal degrees of freedom of the
cluster deposited on a thermalized substrate. The center of mass trajectory is 
integrated using the quaternion algorithm \cite{mddiff,A&T}.
Surprisingly, the diffusion constant  follows the same power law as in the 
free cluster case \cite{mddiff}. This result proves that the diffusion 
mechanism in this 
case cannot be simply explained in terms of dislocation migration within 
the cluster as proposed to explain the diffusion of 2D islands in
\cite{homohamil,hetehamil}. As the substrate atoms are tethered to their 
lattice site, strong elastic deformations or dislocations within the lattice  
are also excluded.  Hence, the motor for diffusion is here the vibrational 
motion of the substrate, and its efficiency appears to be comparable to that 
of the internal cluster modes.

Very recently, U. Landmann performed MD simulations of diffusion of
large gold clusters
on HOPG substrates \cite{isspic9}. He finds high cluster mobility, 
in agreement with
the preceding simulations. His studies show that cluster diffusion in this
case proceeds by two different mechanisms : long (several cluster diameters)
linear "flights" separated by relatively slow diffusive motion as observed
in the preceding simulations. Further work is needed to ascertain the
atomic mechanisms leading to this kind of motion.

\subsubsection{Discussion}

What are the (partial) conclusions which can be drawn from these studies
of cluster diffusion? I think that the main parameter determining the mobility
of 3D islands on a substrate is the possible epitaxy of the cluster
on the substrate. Indeed, if the island reaches an epitaxial orientation,
it is likely to have a mobility limited by the individual atomic movements,
which give a small diffusion constant (of order $10^{-17} cm^2 s^{-1}$
at room temperature). Diffusivities of this magnitude will not affect
the growth of cluster films during typical deposition times, and clusters 
can be considered immobile. The effect of these kind of diffusion rates
can only be seen by annealing the substrates at higher temperatures
or for long times. According to Hamilton \cite{homohamil}, dislocations 
could propagate even for epitaxial islands, but it is likely that
this mechanism is more important in the case of heteroepitaxial islands which I
now proceed to discuss. Indeed, if the island is not in epitaxy on
the substrate, high mobilities can be observed because the cluster sees
a potential profile which is not very different from that seen by a single atom.
It should be noted
that this non-epitaxy can be obtained when the two lattice parameters
(of the substrate and the island) are very different, or also when
they are compatible if there is relative misorientation. The latter has
been observed for gold on ionic substrates \cite{kern} and mobility
is relatively high until the crystallites reach epitaxy. 
The MD simulations presented above show that, for Lennard-Jones 
potentials, only homoepitaxy prevents clusters from moving rapidly on a 
surface. It should be noted that relaxation of the cluster or the 
substrate - which
would favor a locking of the cluster in an energetically favorable 
position at the expense of some elastic energy - has not
been observed in these LJ simulations, nor has dislocation propagation. This
is probably realistic for the low interaction energies which correspond
to metal clusters on graphite. It could also be argued that dislocation 
motion is more difficult in 3D clusters than in 2D islands since the 
upper part of the particle (absent in 2D islands) tends to keep a fixed 
structure. Another important parameter is the 
cluster-substrate interaction : one can think that a large attractive
interaction (for metal on metal systems for example) can induce an epitaxial
orientation and prevent the cluster from diffusing, even in the
heteroepitaxial case. The differences between the diffusion of clusters
grown on a substrate by atom deposition and aggregation and those 
previously formed in a beam
and deposited must also be investigated. One could anticipate that islands
formed by atom aggregation {\it on} the substrate would accommodate easily
to the substrate geometry, whereas preformed clusters
might keep their (metastable) configuration. However, it is 
not at all clear that island nucleation and epitaxy are simultaneous 
phenomena, for it has been observed that islands can form in a somewhat
arbitrary configuration and subsequently orient on the substrate
after diffusion and rotation (see Ref. \cite{kern}).

\subsection{Cluster-Cluster coalescence}
\label{sec_coale}

What happens now when two clusters meet? If they remain simply
juxtaposed, morphologies similar to Fig. \ \ref{sb}a are
observed. In this case, the incident clusters have retained their
original morphology, and the supported particles are identical to
them, even if they are in contact with many others after cluster
diffusion. It is clear, by looking for example to Fig. \ \ref{au} 
that this is not always the case. In these cases, the
supported islands are clearly larger than the incident clusters :
some {\it coalescence} has taken place. How can one understand and predict
the size of the supported particles? Which are the relevant
microscopic parameters? This section tries to answer these
questions, which are of dramatic interest for catalysis, since the
activity of the deposits crucially depends on its specific area and
therefore on the sintering process (See for example Refs. 
\cite{sinter_cata}).

I will first briefly examine the classical theory for sphere-sphere
coalescence (i.e. ignoring the effect of the substrate) and then review 
recent molecular dynamics simulations 
which suggest that this classical theory may not be entirely satisfactory 
for nanoparticles.

\subsubsection{Continuum theory of coalescence}

The standard analysis of kinetics of sintering is due to
Mullins and Nichols \cite{mullins,nichols}. The "motor" of the coalescence is
the diffusion of {\it atoms} of the cluster (or island) surface from the
regions of high curvature (where they have less neighbors and
therefore are less bound) towards the regions of lower curvature. The
precise equation for the atom flux is \cite{nichols}

\begin{equation}
\label{atflux}
\vec{J}_s = -{D_s \gamma \Omega \nu \over k_B T} \overrightarrow{\nabla_sK}
\end{equation}

\noindent 
where $D_s$ is the surface diffusion constant (supposed to
be isotropic), $\gamma$ the surface energy (supposed to be isotropic
too), $\Omega$ the atomic volume, $\nu$ the number of atoms per unit
surface area, $k_B$ Boltzmann's constant, T the temperature and K the
surface curvature (K=1/$R_1$ + 1/$R_2$) where $R_1$ and $R_2$ are the
principal radii of curvature).  For sphere-sphere coalescence, an order 
of magnitude estimation of the shape changes induced by this flux 
is \cite{nichols} :

\begin{equation} \label{dndt}
 {\partial n \over \partial t} \sim 2 B
{\partial^2 K \over \partial s^2} (y=s=0) \end{equation}

where $dn$ is the outward normal distance traveled by a surface
element during $dt$, $s$ the arc length and $B= D_s \gamma \Omega^2
\nu / k_B T$ (the $z$ axis is taken as the axis of revolution). For
this geometry, Eq. \ref{dndt} becomes (Fig. \ \ref{circcoale}) :

\begin{equation}
\label{dxdt}
{\partial l \over \partial t} \sim {B \over l^3} \left(1-{l \over R} \right)
\end{equation}

where I have made an order of magnitude estimation of the second
derivative of the curvature : $\partial K /\partial s \sim
(K(R)-K(l))/l$ and similarly $\partial^2 K /\partial s^2 \sim
(1-l/R)/l^3$ (see Fig. \ \ref{circcoale}).  Integrating Eq. \
\ref{dxdt} leads to

\begin{equation}
\label{xcoalet}
l \sim \left(r^4 + 4Bt \right)^{1/4} \ for \ l \ll R
\end{equation}

Eq. \ \ref{xcoalet} gives an estimation of the coalescence kinetics
for two spheres of radius r and R.

However, despite its plausibility, Eq. \ \ref{xcoalet} has to be used
with care.  First, the calculation leading to it from the expression
of the flux (Eq. \ \ref{atflux}) is only approximate. More
importantly, Eq. \ \ref{atflux} assumes {\it isotropic} surface
tension and diffusion coefficients.  While this approximation may be
fruitful for large particles (in the $\mu$m range, \cite{heyraud}), 
it is clearly wrong for
clusters in the nanometer range. These are generally facetted
\cite{llewis,fluelli,borel} as a result of anisotropic surface
energies. This has two important consequences : first, since the
particles are not spherical, the atoms do not feel a uniform
curvature. For those located on the planar facets, the curvature is
even 0, meaning that they will not tend to move away
spontaneously. This effect should significantly reduce the atomic
flux.  Second, the diffusion is hampered by the edges between the
facets \cite{manninen} which induce a kind of "Schwoebel" effect
\cite{schwoebel}.  Then, the effective mass transfer from one end of
the cluster to the other may be significantly lower than expected from
the isotropic curvatures used in Eq. \ \ref{atflux}.  For these
anisotropic surfaces, a more general formula which takes into account
the dependence of $\gamma$ on the crystallographic orientation should
be used (see for example Ref. \cite{vipi}).  However, this
formula is of limited practical interest for two reasons. First, the
precise dependence of the surface energy on the crystallographic
orientation is difficult to obtain.  Second, as a system of two
touching facetted clusters does not in general show any symmetry, the
solution to the differential equation is hard to find. One possibility
currently explored \cite{misbah} is to assume a simple analytical
equation for the anisotropy of 2D islands and integrate numerically the 
full (anisotropic) Mullins' equations. 

\subsubsection{Molecular Dynamics simulations of coalescence}

Since continuum theories face difficulties in characterizing the evolution
of nanoparticle coalescence, it might be useful to perform molecular dynamics
(MD) studies of this problem. Several studies \cite{llewis,yu,averback} have
been performed, showing that two distinct and generally subsequent 
processes lead to coalescence for particles in the nanometer range : 
plastic deformation \cite{averback} and slow surface diffusion 
\cite{llewis,yu}.

Zhu and Averback \cite{averback} have studied the first stages (up to 160 ps)
of the coalescence of two single-crystal copper nanoparticles (diameter 4.8 nm).
Fig. \ref{cucoale} presents four stages of the coalescence process,
demonstrating that plastic deformation takes place (see the arrows
indicating the sliding planes) and that a relative rotation of the
particles occurs during this plastic deformation (c and d). During the 
first 5 ps, the deformation is elastic, until the elastic limit (roughly
0.8 nm \cite{averback}) is reached : after this, since the shear stress
(Fig. \ref{stress}) is very high, dislocations are formed and glide on (111)
planes in the $<110>$ direction, as usually seen in fcc systems. 
Fig. \ref{stress}
also shows that after 40ps (i.e. Fig. \ref{cucoale}c) the stress on the glide
plane is much smaller and dislocation motion is less important : the
two particles rotate until a low-energy
grain boundary is found (Fig. \ref{cucoale}d). This intial stage of the
coalescence, where the two particles reorient and find a low-energy
configuration, is very rapid, but does not lead in general to 
thorough coalescence. An interesting exception might have been found by Yu and
Duxbury \cite{yu} : their MD simulations showed that for very small 
clusters (typically less than 200 atoms) 
coalescence is abrupt provided the temperature is sufficiently close to the
melting temperature. They argue that this is due to a (not specified)
"nucleation process" : plastic deformation is a tempting possibility.

For larger clusters, the subsequent stages are much slower and imply a 
different mechanism :
atom {\it diffusion} on the surface of the particles. The intial stages
of this diffusion-mediated coalescence have been studied by Lewis et
al. \cite{llewis}. The point was to study if Mullins' (continuum)
predictions were useful in this size domain. In Lewis et al. simulations, 
the embedded-atom method (EAM) \cite{fbd} 
was used to simulate the behavior of {\it unsupported} gold clusters for
relatively long times ($\sim$ 10ns). Evidently, an important role of the
substrate in the actual coalescence of supported clusters is to ensure
thermalization, which is taken care of here by coupling the system to a
fictitious ``thermostat'' \cite{MD}. One therefore expects these coalescence 
events to be relevant to the study of supported clusters in the case where 
they are loosely
bound to the substrate, e.g., gold clusters on a graphite substrate. Strong
interaction of the clusters with the substrate may be complicated and lead 
to cluster deformation even for clusters deposited at low energies, for 
example if the cluster wets the substrate \cite{mdau,hou}. 

Fig. \ \ref{neck} shows the evolution of the ratio $x/R$, where $x$ is 
the radius of the neck between the two particles. After an extremely 
rapid approach of the two
clusters due to the mechanisms studied above (plastic deformation), a slow
relaxation to the spherical shape starts (Fig. \ \ref{coale}). 
The time scale for the slow
sphericization process is difficult to estimate from Fig.\
\ref{neck}, but it would appear to be of the order of a few hundred
ns or more. This number is substantially larger than one would expect
on the basis of phenomenological theories of the coalescence of two
soft spheres. Indeed, Ref. \cite{nichols} predicts
a coalescence time for
two identical spheres $\tau_c = k_BT R^4 /(C D_s \gamma a^4)$, where
$D_s$ is again the surface diffusion constant, $a$ the atomic size,
$\gamma$ the surface energy, $R$ the initial cluster radius, and $C$ a
numerical constant ($C=25$ according to Ref.\ \cite{nichols});
taking $D_s \sim 5 \ 10^{-10}m^2s^{-1}$ (the average value found in
the simulations, see \cite{llewis}), $R=30$ \AA, $\gamma \approx 1 J
m^{-2}$, and $a=3$ \AA, this yields a coalescence time $\tau_c$ of the
order of 40 ns. The same theories, in addition, make definite
predictions on the evolution of the shape of the system with time. In
particular, in the tangent-sphere model, the evolution of the ratio
$x/R$ is found \cite{nichols} to vary as $x/R \sim
(t/\tau_c)^{1/6}$ for values of $x/R$ smaller than the limiting value
$2^{1/3}$. In Fig.\ \ref{neck}, the prediction of this
simple model (full line) is compared with the results of 
the present simulations. There is no agreement between model 
and simulations. The much longer coalescence time observed has been
attributed \cite{llewis} to the presence of facets on the initial 
clusters, which
persist (and rearrange) during coalescence. The facets can be seen in
the initial and intermediate configurations of the system in Fig.\
\ref{coale}; the final configuration of Fig.\ \ref{coale} shows that
the cluster is more spherical (at least from this viewpoint), and that
new facets are forming. That diffusion is slow can in
fact be seen from Fig.\ \ref{coale}: even after 10 ns, at a
temperature which is only about 200 degrees below melting for a
cluster of this size, only very few atoms have managed to diffuse a
significant distance away from the contact region. 

The precise role of the facets in the coalescence process is a subject of
current interest. Experiments have shown that shape evolution is very slow 
in presence of facets for 3D crystallites (see for example \cite{henryfac})
and recent experiments \cite{coale2dexp} and computer simulations 
\cite{coale2d} on 2D islands suggest that the presence of facets can be
effective in slowing down the coalescence process. Clearly, more work is needed
to get a {\it quantitative} understanding of nanoparticles coalescence, and to
evaluate the usefulness of Mullins' approach, especially if one manages
to include the crystalline {\it anisotropy} (see also Refs. \cite{roughening}).

\subsection{Island morphology}

Now I can turn on to the prediction of one of the essential 
characteristics of cluster films : the size of the supported particles. 
As I have already mentioned in the introduction, the size of the
nanoparticles controls many interesting properties of the films.
Therefore, even an approximate result may be useful, and this is
what I obtain in this section.

The experiments shown above demonstrate that the supported particles
can have a variety of sizes, from that of the incident clusters 
($Sb_{2300}$/HOPG, Section \ref{sb2300}) up to many times this size 
(for example $Au_{250}$/HOPG, Section \ref{fhach}). To understand how 
the size of the supported particles is determined, one can look at
a large circular island to which clusters are
arriving by diffusion on the substrate (Fig. \ \ref{circcoale}).
There are two antagonist effects at play here. One is given by
thermodynamics, which commands that the system should try to minimize
its surface (free) energy. Therefore one expects the clusters touching
an island to coalesce with it, leading to compact (spherical)
domains. The other process, driving
the system away from this minimization is the continuous arrival of
clusters on the island edge. This kinetic effect tends to form
ramified islands. What is the result of this competition? Since there 
is a kinetically driven ramification process, it is 
essential to take into account the {\it kinetics} of cluster-cluster
coalescence, as sketched in the previous section. I will use 
Eq. \ \ref{xcoalet} even if it is only approximate, to derive an upper
limit for the size of the compact domains grown by cluster deposition.
It is an {\it upper} limit since, as pointed out in the previous section, 
coalescence for facetted particles could be {\it slower} than predicted 
by Eq. \ \ref{xcoalet}, hence diminishing the actual size of these domains.

We first need an estimate of the kinetics of the second process : the
impinging of clusters on the big island. A very simple argument is
used here (see also \cite{compactatom} for a similar analysis for
atomic growth) : since the number of clusters reaching the surface is
$F$ per unit surface per second and the total number of islands is
$N_t$ per unit surface, each island receives in average a cluster
every $t_r = N_t/F$.

We are now in a position to quantify the degree of coalescence in a
given growth experiment. Let me suppose that a cluster touches a large
island at t=0. If no cluster impinges on the island before this
cluster completely coalesces (in a time $\tau_c$ according to 
Eq. \ \ref{xcoalet}), then
the islands are compact (circular). Instead, if a cluster touches the
previous cluster before its total coalescence has taken place, it will
almost freeze up the coalescence of the previous cluster. The reason is that
now the atoms on the (formerly) outer surface of the first cluster do not feel
curvature since they have neighbors on the second cluster. The
mobile atoms are now those of the second cluster (see
Fig. \ \ref{circcoale}, label a) and the coalescence takes a longer
time to proceed (the atoms are farther form the big island). Then, if $t_r \ll
\tau_c$, the islands formed on the surface are ramified.  For
intermediate cases, the size $R_c$ of the compact domains can be
estimated from Eq. \ref{xcoalet} as $R_c = x(\tilde{t}_r)$ 
where $\tilde{t}_r$ takes into
account the fact that, to freeze the coalescence of a previous
cluster, one cluster has to touch the island at roughly the same point
: $\tilde{t}_r \simeq t_r 2\pi R/r$ and

\begin{equation}
\label{Rcsub}
{R_c}^4 = r^4 + 4B{2\pi R_c \over r} {N_t \over F}
\end{equation}

Eq. \ \ref{Rcsub} describes the limiting cases ($B \sim \infty$  or 
$B \sim 0$) correctly. The problem with the intermediate cases is
to obtain a reliable estimate of the (average) atomic surface self diffusion. For gold, Chang and Thiel \cite{thiel}
give values which vary between 0.02 eV on compact facets and 0.8 eV on 
more open surfaces. One solution is to go the other way round and estimate
$D_s$ from the experimental data and Eq. \ \ref{Rcsub}. From Fig. \
\ref{au}, estimating $R_c$ from the thickness of the island arms, and
using the experimental values for r (0.85 nm) and the fact that since
the flux is pulsed (see Section \ref{fhach}), the time between two successive
arrivals of clusters is approximately the time between two pulses (0.1
s), and not $N_t/F$, one obtains $D_s \simeq 3 \ 10^{-3} cm^2 s^{-1}
exp(-0.69 eV/(k_B T))$, which seems a sensible value.

Despite the difficulty of defining average diffusion coefficients, one can
use Eq. \ \ref{Rcsub} to obtain a reasonable guess for the size of the 
compact domains by assuming that $D_s$ is thermally activated : 
$D_s(T)=D_0 exp(-E_a/(k_B T))$ with a prefactor $D_0 = 10^{-3} cm^2 s^{-1}$ 
and an 
activation energy $E_a$ taken as a fraction of the bonding energy between 
atoms (proportional to $k_BT_f$). One obtains \cite{moi_unpub}

\begin{equation}
\label{BT}
B=10^{11} exp(-4.6 T_f/T) nm^4/s
\end{equation}

Inserting this value in Eq. \ \ref{Rcsub} leads to Fig. \ \ref{RcT}
where the size of the compact domains is plotted as a function of
$T/T_f$. The important feature is that as long as $T/T_f \leq 1/4$,
the incident particles do not merge. Note that this $1/4$ is sensitive
to the assumed value of $D^*$, but only via its logarithm. Again, this
estimation of $R_c/r$ is an upper limit since coalescence could
be slower than predicted by Eq. \ \ref{xcoalet}.

\vspace{.5cm}

What would happen now if the incident clusters were {\it liquid}?  An
experimental example of this liquid coalescence is given by the
deposition of $In_{100}$ on a-C (see above).  A rough guess of the
coalescence time is given by a hydrodynamics argument \cite{barrat}:
the driving force of the deformation is the surface curvature $\gamma
/ R^2$ where $\gamma$ is the liquid surface tension and $R$ the
cluster radius. This creates a velocity field which one can estimate
using the Navier-Stokes equation : $\eta \Delta v = \gamma / R^2$
where $\eta$ is the viscosity and v is the velocity of the fluid.  This 
leads to $\tau_c(liquid)
\sim R/v \sim \eta R/\gamma$. Inserting reasonable values for both
$\eta$ (.01 Pa s) and $\gamma$ (1 J $m^{-2}$) leads to
$\tau_c(liquid) \sim 0.01 R$ which gives $\tau_c(liquid) = 10 ps$ for
$R \sim 1nm$. This is the good order of magnitude of the
coalescence times found in simulations of liquid gold clusters
($\tau_c(liquid) \sim 80 ps$ \cite{llewis}).  Now, since
$\tau_c(liquid) \ll t_r$ ($t_r \sim 0.1 s$, see above) cluster-cluster
coalescence is almost instantaneous, which would lead to $R_c \sim
\infty$.  In fact, $R_c$ is limited in this case by {\it static}
coalescence between the big islands formed during the growth. The
reason is that the big islands may be solid or pinned by defects
leading to a slow coalescence. The analysis is similar here to what
has been done for atomic deposition \cite{jeffers}.

\subsection{Thick films}

The preceding section has studied the first stages of the growth, the
submonolayer regime, which interests researchers trying to
build nanostructures on the surface. I attempt here a shorter study of
the growth of thick films, which are known to be very different from
the bulk material in some cases \cite{siegel_ency,nanomat_revue,LECD}. 
The main reason for this
is their nanostructuration, as a random stacking of nanometer size
crystallites. Therefore, it is interesting to understand how the size
of these crystallites is determined and how stable the nanostructured film is.
One can anticipate that the physical mechanism for cluster size evolution 
is, as in the submonolayer case, sintering by atomic diffusion. For thick
films however, surface diffusion can only be effective before a given cluster
has been "buried" by the subsequent deposited clusters. Thus, most of
the size evolution takes place during growth, for after the physical
routes to coalescence (bulk or grain boundary diffusion) are expected
to be much slower. Studies of compacted nanopowders \cite{siegel_ency,nieman}
have shown that nanoparticles are very stable against grain growth. Siegel 
 \cite{siegel_ency} explains this phenomenon in the following way.
The two factors affecting the chemical potential of the atoms, and
potentially leading to structure evolutions are local differences in
cluster size or in curvature. However, for
the relatively uniform grain size distributions and flat grain boundaries
observed for cluster assembled materials \cite{siegel_ency}, these two
factors are not active, and there is nothing telling locally to the atoms 
in which direction to migrate to reduce the global energy.  Therefore,
the whole structure is likely to be in a deep local (metastable) minimum 
in energy, as
observed in closed-cell foams. The stability of such structures has been
confirmed by several computer simulations \cite{zhu,celino} which have
indicated a possible mechanism of grain growth at very high ($T/T_f \sim 0.8$)
temperatures : grain boundary amorphization or melting \cite{zhu}.

What determines the size of the supported particles {\it during} the growth?
For thick films, a reasonable assumption is that a cluster impinging
on a surface already covered by layer of clusters does {\it not}
diffuse, because it forms strong bonds with the layer of the deposited
clusters. This hypothesis has been checked for the growth of
$Sb_{2300}$ on graphite \cite{ss}. There are two main
differences with submonolayer growth : first, an impinging cluster 
has more than one neighbor
and the sphere-sphere kinetics is not very realistic; second the
relevant time for ramification I have used in the preceding section
is no more useful here since clusters do not move.  As a first
approximation, to obtain an upper limit in the size of the domains, we
can use the same coalescence kinetics and take a different
"ramification" time : the average time for the arrival of a cluster
touching another is roughly $t_f \sim 1/(Fd^2)$ where $d=2r$ is the
diameter of the cluster. If the same formula (Eq. \ \ref{xcoalet}) is 
used, one finds

\begin{equation}
\label{Rcthick}
{R_c}^4 = r^4 + {B \over Fr^2}
\end{equation}

The results obtained using the same approximation as in the preceding
section for $B$ (Eq. \ref{BT}) are shown in Fig. \ref{RcT}. 

Experimentally, there are observations for deposition of Ni and Co
clusters \cite{juliette}. The size of the crystallites is comparable 
to the size of the incident (free) clusters. This is compatible
with Eq. \ref{Rcthick} since the $T_f$ of these elements is
very high ($\simeq 1800 K$). Therefore, Eq. \ \ref{Rcthick} predicts that
films grown at $T=300K$, ($T/T_f \sim 0.17$) should keep a 
nanostructuration with $R_c \simeq r$, as is observed experimentally 
\cite{juliette}. I stress again 
that a structure obtained with cluster deposition with this
characteristic size is not likely to recrystallize in the bulk phase
(thereby loosing its nanophase properties) unless brought to
temperatures close to $T_f$ \cite{siegel_ency,zhu,celino}.

\section{CONCLUSIONS, PERSPECTIVES}

What are the principal ideas presented in this paper? 

First, useful models to analyze the first stages of thin film growth
by cluster deposition have been presented in detail (Section \ref{sec_models}).
These models are useful at
a fundamental level, and I have shown in Section \ref{exps} how 
many experimental results concerning
{\it submonolayer} growth can be interpreted by combining these few 
simple processes (deposition, diffusion, evaporation \ldots).
Specifically, by comparing the experimental evolution of the island 
density as a function of the number of deposited particles to the predictions
of computer simulations, one can obtain {\it quantitative} information
about the relevant elementary processes. 

Second, the quantitative
information on diffusion has shown that large clusters can move rapidly
on the surface, with diffusion constants comparable to the atomic ones.
A first attempt to understand this high diffusivity at the atomic level
is given in Section \ref{sec_atomic} :
the conclusion is that rapid cluster diffusion might be quite common, 
provided the cluster and the substrate do not find
an epitaxial arrangement.  Concerning cluster-cluster coalescence, it has
been suggested that this process can be much slower than predicted by the 
usual sintering theories \cite{nichols}, probably because 
of the cluster {\it facets}.

Third, despite all the approximations involved in its derivation, 
Fig. \ \ref{RcT} gives an important information on the morphology of the 
film : an {\it upper} limit for the ratio of the size of the compact 
domains over the size of the incident clusters. This helps
understanding why cluster deposition leads to nanostructured films provided 
the deposition temperature is low compared to the fusion temperature of
the material deposited ($T_s \leq T_f/4$). Clearly, further experimental 
and theoretical work is needed in order to confirm (or invalidate) 
Fig. \ \ref{RcT}.

Ii is clear that we still need to understand many aspects of the physics
of cluster deposition. Possible investigation directions include the
following, given in an arbitrary order. First, the coalescence of 
nanoparticles has yet to be understood
and quantified. This is a basic question for both submonolayer and
thick materials. Second, one has to characterize better the interactions
between clusters and the substrate, and especially its influence on
the cluster diffusion. It is also important to investigate the possible
interactions between the clusters, which could dramatically affect the
growth. Obtaining {\it ordered} arrays of nanoparticles is a hot topic
at this moment. A possibility is the pinning of the clusters on 
surface "defects" which demands a better understanding 
of cluster interaction with them. Another idea is to use the self-organization
of some bacteria to produce an ordered array on which one
could arrange the clusters (See Ref. \cite{biolo}, especially Chap. 5). 
Clearly, investigating the interaction
of clusters with biological substrates is not an easy task, but it is
known that practical results are not always linked to a clear understanding of
the underlying mechanisms \ldots

\vspace{2cm}

{\it Acknowledgments} : This article could never be written without
all the experimental and theoretical work carried out in our group in
Lyon and in collaboration with other groups. On the experimental side,
the Center for the study of small clusters ("Centre pour l'\'etude des
petits agr\'egats") gathers researchers from solid state physics
("D\'epartement de Physique des Mat\'eriaux", DPM), the gas phase
("Laboratoire de Spectrom\'etrie Ionique et Mol\'eculaire", LASIM) and
catalysts ("Institut de Recherche sur la Catalyse", IRC). I therefore
acknowledge all their researchers for their help, and especially those
who have done part of the research presented here : Laurent Bardotti,
Michel Broyer, Bernard Cabaud, Francisco J. Cadete Santos Aires, V\'eronique
Dupuis, Alain Hoareau, Michel Pellarin, Brigitte Pr\'evel, Alain
Perez, Michel Treilleux and Juliette Tuaillon. On the theoretical
side, this work was carried out in collaboration with Jean-Louis
Barrat, Pierre Deltour and Muriel Meunier (DPM, Lyon), my friend Hern\'an
Larralde (Instituto de F\'{\i}sica de Cuernavaca, Mexico), Laurent
Lewis (Universit\'e de Montr\'eal, Canada) and Alberto Pimpinelli
(Universit\'e Blaise Pascal Clermont-2, France). I am happy to thank 
Claude Henry (CRMC2, Marseille) and Horia Metiu (University of California)
for a careful reading of the manuscript,
Jean-Jacques M\'etois (CRMC2, Marseille) for interesting discussions, 
Michael Moseler (Freiburg University) for sending me Fig. \ \ref{md_eci}
and Simon Carroll for Fig. \ \ref{palmerag}.

{\it $^*$} e-mail address : jensen@dpm.univ-lyon1.fr

\begin{figure}
\caption{
{\it Ag nanoislands grown on two monolayers of Ag deposited on Pt(111) and 
annealed at 800K. The inset shows a fast fourier transform of the spatial
distribution. From Ref. \protect\cite{revharri})}
}
\label{triangles}
\end{figure} 

\begin{figure}
\caption{
{\it By changing the mean size of the incident antimony clusters, one can
dramatically change the morphology of the submonolayer film. The four
micrographs have been obtained for the same thickness (1 nm) and deposition
rate ($5 \times 10^{-3} nm s^{-1}$). The mean sizes are : (a) $Sb_4$, 
(b) $Sb_16$, (a) $Sb_36$, (a) $Sb_240$. The changes
in morphology are interpreted by the different mobilities of the clusters
as a function of their size, as well as their different coalescence dynamics
and sensitivity to surface defects. From Ref. \protect\cite{fuchs}.} }
\label{sbfuchs}
\end{figure} 

\begin{figure}
\caption{
{\it Molecular Dynamics simulations of the morphology of films 
obtained by $Mo_{1043}$ cluster deposition for increasing incident kinetic energies per atom (as indicated in the figures) onto a Mo(001) substrate. 
From Ref. \protect\cite{eci_md}}
}
\label{md_eci}
\end{figure} 

\begin{figure}
\caption{
{\it Principle of a Kinetic Monte Carlo simulation (see text).} }
\label{kmccategories}
\end{figure} 

\begin{figure}
\caption
{\it Main elementary processes considered in this paper for the
growth of films by cluster deposition. (a) adsorption of a cluster
by deposition; (b) and (d) diffusion of the isolated clusters on the 
substrate; (c) formation of an island of two monomers by juxtaposition
of two monomers (nucleation) (d) growth of a supported island by 
incorporation of a diffusing cluster (e) evaporation of an adsorbed
cluster. I also briefly consider the influence of island diffusion (f).}
\label{dda}
\end{figure} 

\begin{figure}
\caption{
{\it Possible interaction of two clusters touching on the surface :
(a) pure juxtaposition (b) total coalescence. Intermediate cases (partial
coalescence) are possible and will be described later.}
}
\label{interaction}
\end{figure}

\begin{figure}
\caption
{\it Time scales of some elementary processes considered in this paper 
for the growth of films by cluster deposition. The relevant processes
are those whose time scale are smaller than the deposition time scale
shown by the arrow in the left. In this case, models including only
cluster diffusion on the substrate and cluster cluster coalescence 
are appropriate. "Island diffusion" refers to the motion of islands of clusters
as a whole, "cluster dissociation" to the evaporation of atoms {\it from} the
cluster and "interdiffusion" to the exchange of atoms in the cluster with
substrate atoms.}
\label{timescale}
\end{figure}

\begin{figure}
\caption{
{\it Evolution of the monomer and island densities as a function of
the thickness (in monolayers), for islands formed by pure
juxtaposition : (a) complete condensation, $F=10^{-8}, \
\tau_e=10^{10} \ (\tau=1)$. These values mean : $X_S = 10^5$ and
$\ell_{CC} = 22$ (b) important evaporation, $F=10^{-8}, \ \tau_e=600 \
(\tau=1)$ ($X_S = 25$ and $\ell_{CC} = 22$). $\ell_{CC}$ represents
the mean island separation at saturation for the given fluxes when there is
no evaporation \protect\cite{evaprb}.  The length units correspond to
the incident cluster (monomer) diameter.  In (b) the "condensation" curve
represents the total number of particles actually present on the
surface divided by the total number of particles sent on the surface
(F t). It would be 1 for the complete condensation case, neglecting
the monomers that are deposited on top of the islands. The solid line
represents the constant value expected for the monomer concentration,
while the dashed line corresponds to the {\it linear} increase of the
island density (see text).}}
\label{dent} 
\end{figure}

\begin{figure}
\caption{
{\it Morphology of a submonolayer deposit in the case of growth with
complete condensation and pure juxtaposition : (a) $\theta = 0.1$\% ; 
(b) $\theta = 15$\%. The
values of the parameters are : $F=10^{-8} ML/s$, $\tau=1$ and L=300.}
}
\label{morpho}
\end{figure} 

\begin{figure}
\caption{
Saturation island density as a function of the normalized flux ($\tau=1$) for
different growth hypothesis indicated on the figure, always in the
case of island growth by pure juxtaposition. {\it "no evap"} (circles) means
complete condensation. Triangles show the densities
obtained if there is evaporation, for $\tau_e=100$. In the 
preceding cases, islands are
supposed to be immobile. This hypothesis is relaxed for the last set
of data, {\it mobile islands} (squares) , where island mobility is supposed to
decrease as the inverse island size \protect\cite{boston} (there is
no evaporation).  The dashed
line is an extrapolation of the data for the low normalized fluxes. Fits
of the different curves in the low-flux region give : {\it "no evap"} (solid
line): $N_{sat} = 0.53 (F \tau)^{0.36}$ ;  {\it "evap"} (dotted line) : 
$N_{sat} = 0.26 F^{0.67} \tau^{-1/3} \tau_e$ (for
the $\tau$ and $\tau_e$ exponents, see \protect\cite{evaprb} and Appendix A) 
and {\it "mobile islands"} (dashed line) : $N_{sat} = 0.33 (F \tau)^{0.42}$}
\label{nmax2d}
\end{figure} 

\begin{figure}
\caption{Evolution of the island density as a function of the thickness
 for different growth hypothesis. This 
figure shows that the {\it same} saturation density can be obtained for
films grown in very different conditions. The 
different sets of data represent :
{\it triangles} : growth with coalescence and evaporation,
 $\tau_e = 100 \tau$ and $F\tau=1.2 \ 10^{-8}$, {\it circles} : growth
with coalescence but without evaporation ($F\tau=3 \ 10^{-10}$), {\it
solid line} : growth with pure juxtaposition without evaporation
($F\tau=2.5 \ 10^{-9}$), {\it squares} : growth with coalescence on
defects (defect concentration : $5 \ 10^{-4}$ per site) and $F\tau=
10^{-14}$ (no evaporation), {\it dashed line} : growth with pure
juxtaposition without evaporation but with mobile islands,
$F\tau=10^{-8}$.}
\label{cinetns}
\end{figure} 

\begin{figure}
\caption{
Normalized island size distributions for $F=10^{-8}$, $\tau=1$ and different
values of $\tau_e$ for islands formed by pure juxtaposition (no
coalescence). The size distributions were obtained for different
coverages $\theta$ between .05 and 0.2. The solid line shows the size
distribution obtained without evaporation, the dashed line that
obtained with mobile dimers and the numbers show the different values
of $\tau_e$. The size distributions shown here have been obtained with
$F \tau=10^{-8}$, but the same distributions are obtained if both $F$,
$\tau$ and $\tau_e$ are changed but the same parameter $\epsilon =
{\left(1+X_S\right)}X_S^{5}\left(F\tau\right)$ is obtained (see
Ref. \protect\cite{evaprb})}
\label{distrib2d}
\end{figure} 

\begin{figure}
\caption{
Saturation island density as a function of the normalized flux ($\tau=1$) 
for different growth hypothesis in the case of growth by total coalescence
(3d islands). 
I show the densities obtained for the complete condensation case (filled 
circles) and for two different evaporation times :
$\tau_e = 100$ (triangles) and $\tau_e = 20$ (squares). The label {\it defects} 
means growth in presence of defects which act as
nucleation centers. Their concentration is $10^{-3}$ per site.
The dashed line is an extrapolation of the defect data for the low 
normalized fluxes. Fitting the simulation data leads
to the following numerical relations : $N_{sat} = 0.27 (F \tau)^{0.286}$
when there is no evaporation (solid line) ; 
$N_{sat} \sim 0.039 F^{0.55} \tau^{-2/3}\tau_e^{4/3}$ when evaporation is
significant (from an approximation for the two dotted curves) : the exponents 
for $\tau$ and $\tau_e$ have been derived
from a rate-equations treatment (Appendix B)}
\label{nmax3d}
\end{figure}

\begin{figure}
\caption{
Normalized island size distributions obtained for $F=10^{-8}$, $\tau=1$ and
different values of the evaporation time $\tau_e$ for islands formed
by total coalescence. The size distributions were obtained for
different coverages $\theta$ between .05 and 0.2. The solid line shows
the size distribution obtained without evaporation.  The number next
to each symbol corresponds to $\tau_e$.}
\label{distrib3d}
\end{figure} 

\begin{figure}
\caption{
Effect of the presence of defects on the island size distribution. The
rescaled island size distributions are obtained for $F=10^{-8}$ and
different values of the evaporation time $\tau_e$ ($\tau =1$) for
islands formed by total coalescence by nucleation on defects. The size
distributions were obtained for different coverages $\theta$ between
.05 and 0.15. Contrary to what is observed for homogeneous nucleation,
(i.e. without defects) the histograms do depend on the coverage 
for nucleation on
defects. The solid line shows the size distribution obtained without
evaporation.}
\label{distribdef}
\end{figure}

\begin{figure}
\caption{
{\it Values of (a) the thickness $e_{sat}$ and 
(b) island density $N_{sat}$ at the {\it saturation} of island
density as a function of the evaporation parameter $\eta = F \tau
X_S^6$ for growth with pure juxtaposition \protect\cite{evaprb}. The
solid lines represent theoretical predictions \protect\cite{evaprb}.}}
\label{coll2d}
\end{figure} 

\begin{figure}
\caption{
{\it Values of the thickness  $e_{sat}$ (a) and the condensation coefficient $C_{sat}$  (b) at the {\it saturation} of island density in
 the total coalescence limit (3d growth for atomic deposition). In 
the limit of low island densities (i.e. high evaporation rates),
$C_{sat}$ is a {\it constant} (see Ref. \protect\cite{pyr}, this
regime is indicated by the solid line).  However, there are crossover
regimes which depend on the precise $\tau_e$ and which are shown
here. Then, from a measure of $C_{sat}$ and $N_{sat}$ one can get an
estimate for $\tau_e$ for the not too low island densities which
correspond to many experimental cases. In the same spirit, (a) shows
the evolution of $e_{sat}$ as a function of $N_{sat}$ in the crossover
regime. The numbers correspond to the different $\tau_e/\tau$ used for
the simulations and CC refers to the case of complete condensation 
(no evaporation). The dotted line in the higher left shows the limiting 
regime $e_{sat} \sim {N_{sat}}^{-1/2}$.}  }
\label{coll3d}
\end{figure} 

\begin{figure}
\caption{
{\it Typical island morphologies obtained experimentally by TEM 
(a) and from the computer simulations (b)
at the same coverage. (a) $Sb_{2300}$ deposition on graphite HOPG at
$T_s=353K$ and $f=6 10 ^{-3} nm s^{-1}$, corresponding
to $F=1.7 10 ^{-3} ML/s$. The deposited thickness is 0.5 nm or
$e= 0.14ML$ (b) model including only deposition, diffusion and
pure juxtaposition of the incident clusters, $F\tau=9 \ 10^{-11}$}}
\label{sb}
\end{figure} 

\begin{figure}
\caption{
{\it (a) Evolution of the island density as a function of the
deposited thickness. The solid line is a fit to the experimental data
with $F\tau=1.75 \ 10^{-8}$.  (b) Evolution of the maximum island
density (N$_{sat}$) as a function of the incident flux F at room
temperature. The solid line is a fit to the experimental data : we
find N$_{sat}$= a f $^{0,37\pm0.03}$ (c) Dependence of the diffusion
coefficient on the temperature. From a fit on the experimental data
(solid line), one finds D = D$_{0 }$ exp (-E$_a$/kT), with E$_{a }$ =
0.7$\pm$0.1 eV and $D_{0 } = 10^4 cm^2 s^{-1}$}. The island
densities are expressed {\it per site}, a site occupying the projected
surface of a cluster, equivalent to $2.08 \ 10^{-13} cm^2$}
\label{sbss}
\end{figure} 

\begin{figure}
\caption{
{\it Scanning electron microscopy of a submonolayer deposit of $Ag_{160}$
slightly accelerated (50eV) clusters on HOPG. 
From Ref. \protect\cite{palmerdepo}.
}
}
\label{palmerag}
\end{figure} 

\begin{figure}
\caption{
{\it (a) Morphology of a $Sb_{36}$ film at $e=1.8 ML$.
 (b) Evolution of the island density (per site) as a function of
thickness (ML). The dashed line represents a fit of the data with
$F\tau=10^{-5}$ assuming a pyramidal (half-sphere) shape for the
supported islands, while the solid line assumes that islands are
spherical and $F\tau=3 \ 10^{-6}$} }
\label{sb36}
\end{figure} 

\begin{figure}
\caption{
{\it Morphologies of a $Au_{250}$ films at $e=0.12$ ML and increasing
temperatures as indicated in the micrographs. There are less and less
islands as the substrate
temperature is raised and the islands become more and more compact.}
}
\label{au}
\end{figure} 

\begin{figure}
\caption{
{\it (a) Saturation island density as a function of the diffusion time
$\tau$ ($F_i$=6 ML/s) for two hypothesis : only the monomers move 
(solid line) or
islands up to the pentamer move too (dashed line). The lowest island
densities have been extrapolated. (b) Temperature dependence of
diffusion coefficient as derived from (a) and Fig. \protect\ref{au} in the 
two hypothesis : only
the monomers move (solid line) or islands up to the pentamer move too
(dashed line).}}
\label{denhache}
\end{figure} 

\begin{figure}
\caption{
{\it Morphology (a) and diffraction pattern (b) of a $Au_{250}$ 
submonolayer deposit on $NaCl$
at $e=0.12$ ML. The supported islands are small (mean diameter $\simeq 5 nm$)
and in epitaxy with the substrate as shown by the diffraction pattern.}}
\label{aunaclmorph}
\end{figure} 

\begin{figure}
\caption{
{\it Radial distribution functions for gold clusters grown at 293K by atomic
deposition before and after annealing at 390 K for several minutes. The
right inset shows the size distribution of the clusters, which does not
change, demonstrating that no particle-particle coalescence via atomic
evaporation takes place. From Ref. \protect\cite{zanghi}.} }
\label{radial}
\end{figure}

\begin{figure}
\caption{
{\it Individual atomic mechanisms leading to island diffusion. PD refers
to diffusion of atoms on the periphery of the island, while the exchange
of atoms between the island and the atomic 2D gas is shown by the
condensation and evaporation labels. The dashed circles represent the
old positions of the atoms, while the continuous circles represent the
new positions, after the elementary process.} }
\label{PD}
\end{figure}

\begin{figure}
\caption{
{\it Principle of island motion by dislocation propagation. The atomic column
in the middle of the island jumps from fcc to hcp sites, moving the island
center of mass. From  Ref. \protect\cite{homohamil}.} }
\label{disloc}
\end{figure} 

\begin{figure}
\caption{
{\it Successive positions of a $Ir_{19}$ 2D cluster on a Ir(111) surface
observed by field ion microscopy at low temperature. The motion takes 
place at $T \sim 690 K$ and the figures correspond to 6,10 and 14 heating
intervals of 10 seconds each. From  Ref. \protect\cite{ehrlich}.} }
\label{ir19}
\end{figure} 

\begin{figure}
\caption{
{\it Configuration of the Lennard-Jones cluster on the crystalline
surface.  (A) Side view : The cluster is partially wetting the
surface. (B) Top view : The two halves of the cluster have been
colored at the beginning of the run. After the cluster center of mass
has moved by roughly three substrate lattice constants from its
original position, the two parts of the cluster are still well
distinct. Then, the cluster diffusion cannot been explained in terms
of single atom mechanisms (n=100, $\sigma_{ss}$=0.7,
$\epsilon_{sc}$=0.4, T=0.3) } }
\label{mdclu}
\end{figure} 

\begin{figure}
\caption{
{\it (a) Dependence of the diffusion coefficient on the mismatch
between the lattice parameter of the substrate and the cluster. A
small change in the lattice parameter of the cluster leads to a huge
change in the diffusivity (n=100, $\epsilon_{sc}$=0.4, T=0.3, Run
Length = 12500 $\tau$) (b) Dependence of the diffusion coefficient of
a cluster as a function of its number of atoms.  Data correspond
to different mismatches between the cluster and the substrate lattice
parameters.  The diffusion coefficient decreases as a power law with
exponent $\alpha$.  The two different slopes correspond to different
diffusion regimes : the weaker dependence corresponds to a brownian
trajectory; the stronger correspond to a ``hopping-like'' diffusion.
For comparison, the arrow indicates the diffusion coefficient of a
single adatom with $\sigma_{ss}=0.9$.} } 
\label{simudif} 
\end{figure}

\begin{figure}
\caption{
{\it Trajectory of a cluster center of mass diffusing on a substrate. The
solid line represents the trajectory and the circles the equilibrium
position of the surface atoms.
(a) large mismatch : the motion is "brownian-like", i.e. the cluster does
not "see" the structure of the surface. The values of the parameters are : 
$\epsilon_{sc}=0.4$, $\sigma_{ss}=0.7$, T=0.3, n=100;  (b) small mismatch : 
the cluster center of mass jumps from one hexagon center to a nearest 
neighbor one. The values of the parameters are the same as for (a)
except for $\sigma_{ss}=0.9$} } 
\label{trajectory} 
\end{figure}

\begin{figure}
\caption{
{\it Schematic illustration of the competition between coalescence and
kinetic ramification. $R$ is the radius of the largest island, $r$
that of the incident clusters and $l$ stands for the typical length of
a coalescing cluster. The label "a" refers to the ramification
process when a cluster is touched by another one before coalescence
can take place.}  }
\label{circcoale}
\end{figure} 

\begin{figure}
\caption{
{\it Atomic positions at four different times during 
partial coalescence of Cu clusters (4.8 nm diameter each). The atomic
positions are projected onto the (12$\bar1$) plane of the bottom sphere
for (a) to (c) and onto the (10$\bar1$) plane for (d). The arrows 
indicate the sliding plane ((a) to (c)) and the grain-boundary 
dislocation (d). From Ref. \protect\cite{averback}.}}
\label{cucoale}
\end{figure} 

\begin{figure}
\caption{
{\it Distribution of shear stress in the sliding plane at
different times during the coalescence shown in Fig. \protect\ref{cucoale}.
Filled circles : 5ps, open squares : 10ps, open triangles : 20ps, crosses : 30ps
and plus : 40ps. From Ref. \protect\cite{averback}.}}
\label{stress}
\end{figure} 

\begin{figure}
\caption{
{\it Evolution in time of the ratio of the neck radius, $x$, to the cluster
radius, $R$. The full line represents the numerical solution obtained by
Nichols \protect\cite{nichols} with an arbitrary time scale, while the 
crosses are the results of Lewis et al. \protect\cite{llewis} simulations.
}
}
\label{neck}
\end{figure} 

\begin{figure}
\caption{
{\it Successive cluster morphologies during the coalescence of a gold
767-atom liquid cluster with another gold 1505-atom 
solid cluster. The figures represent three stages of the 
coalescence process after 0, 1 and 10 ns, i.e. times much longer
than those studied in Ref. \protect\cite{averback}.}}
\label{coale}
\end{figure} 

\begin{figure}
\caption{
{\it Approximate dependence of the radius of the supported particles
$R_c$ as a function of the substrate temperature for submonolayer and thick
films. Lines refer to predictions
from Eq. \ref{Rcsub} with different incident cluster radius while symbols
represent experimental results shown. The theoretical predictions for
the submonolayer regime were
obtained by taking $N_t/F=0.1$ and using Eq. \ref{BT}. For the thick
film limit, I have taken $F=10^{-3} ML/s$ and $r=2.5nm$ (Eq. \ref{Rcthick}). 
One should consider
these theoretical $R_c$ values as an {\em upper} limit since coalescence
may be much slower at these (nano)scales (see the text). As a consequence,
it is no surprise that the predicted values
are clearly larger than the experimental ones.
Concerning Sb, the huge difference can come from a partial oxidation of
the clusters on the substrate because of the relatively bad vacuum conditions
(pressure $\sim 10^{-7}$ Torr). Even a thin oxide layer can decrease
significantly atomic surface diffusion and transport, thus slowing 
the coalescence process.} } 
\label{RcT} 
\end{figure}

\newpage

\begin{center}
{\bf Appendix A}

Regimes predicted by rate-equations calculations for the growth of 2D 
islands with evaporation. These predictions agree with the computer
simulations presented in this paper and are relevant for both cluster
and atomic deposition (see Ref. \protect\cite{evaprb} 
for more details).
\end{center}

\vspace{2cm}

I will here rapidly recall how the rate-equations can be written 
\protect\cite{evaprb}, and then turn on to the different regimes
which can be derived from them.

The rate equation describing the time evolution of the
density $\rho$ of monomers on the surface is, to lowest relevant
orders in F:
\begin{equation}
{d\rho\over dt}=F(1-\theta)-{\rho\over\tau_e} -F\rho - 
2\sigma_o\rho -\sigma_i N_t
\label{rho}
\end{equation}
The first term on the right hand size denotes the flux of monomers
onto the island free surface, ($\theta$ is the island coverage
discussed below).  The second term represents the effect of
evaporation, i.e. monomers evaporate after an average time
$\tau_e$. The third term is due to the possibility of losing monomers
by effect of direct impingement of a deposited monomers right beside a
monomer still on the surface to form an island. This ``direct impingement'' 
term is usually negligible, and indeed
will turn out to be very small in this particular equation, but the
effect of direct impingement plays a crucial role in the kinetics of
the system in the high evaporation regimes. The last two terms
represent the loss of monomers by aggregation with other monomers and
with islands respectively. The factors $\sigma_o$ and $\sigma_i$ are
the ``cross sections'' for encounters and are calculated in Refs.
\cite{venables73,bales,evaprb}.

The number $N_t$ of islands will be given by:
\begin{equation}
{dN_t \over dt}=F\rho+\sigma_o\rho
\label{islands}
\end{equation}
where the first term represents the formation of islands due to direct
impingement of deposited monomers next to monomers already on the
surface, and the second term accounts for the formation of islands by
the encounter of monomers diffusing on the surface.

For the island coverage $\theta$ i.e. the area covered by all the
islands per unit area, one has:
\begin{equation}
{d\theta\over dt}=2\left[F\rho+\sigma_o\rho\right]
                  +\sigma_i N_t + JN_t 
\label{cove}
\end{equation}

The term in brackets represents the increase of coverage due to
formation of islands of size 2 (i.e. formed by two monomers) either by
direct impingement or by monomer-monomer aggregation. The next term
gives the increase of coverage due to the growth of the islands as a
result of monomers aggregating onto them by diffusion, and the last
term represents the growth of the islands due to direct impingement of
deposited monomers onto their boundary, or directly on the island. This
last term depends on $X_S^*$, the desorption length of
monomers diffusing on top of the islands \cite{evaprb}. In all the
simulations presented here (Section \ref{sec_models}), I have taken
$X_S^* = 0$. The 
total surface coverage is given by $\theta+\rho\sim\theta$ except at 
very short times.

The cross sections can be evaluated in the quasistatic approximation,
which consists in assuming that $R$ does not vary in time and that the
system is at a steady state. One finds \cite{venables73,bales,evaprb}

\begin{equation}
\sigma_i=2\pi RD\left({dP\over dr}\right)_{r=R}=2\pi D\rho 
\left(R\over{X_S}\right){K_1(R/X_S)\over K_0(R/X_S)}
\end{equation}
The cross section for monomer-monomer encounters $\sigma_o$ is
obtained from the same formula substituting $R$ by the monomer radius,
and $D$ by $2D$ as corresponds to relative diffusion. 

After some additional approximations, one finds \cite{evaprb}
three principal regimes which are spanned as the
evaporation time $\tau_e$ decreases. They have been called :
{\it complete condensation} regime where evaporation is not important,
{\it diffusion} regime where islands grow mainly by diffusive
capture of monomers and finally {\it direct impingement} regime
where evaporation is so important that islands can grow only by
capturing monomers directly from the vapor.  Within each of these
regimes, there are several subregimes characterized by the value of
$X_S^*$. I use $\l_{CC} \equiv
(F \tau)^{-1/6}$, the island-island distance at saturation when there is no
evaporation and $R_{sat}$ as the maximum island radius, reached
at the onset of coalescence. 

\vspace{3cm}

{\bf complete \ condensation } $X_S \gg \l_{CC}$

\begin{equation}
N_{sat}\sim F^{1/3}\tau^{1/3} {\rm \ for \ any \ X_S^*}
\label{eqcc}
\end{equation}

\vspace{1cm}

{\bf diffusive growth } $1 \ll X_S \ll \l_{CC}$

\begin{equation}
N_{sat}\sim\cases{(FX_S^2\tau_e)^{2/3}(X_S+X_S^*)^{-2/3} & {\rm if}~~~~
                   $X_S^* \ll R_{sat}$ \  (a)\cr\cr
                   F\tau_eX_S^2 & {\rm if}~~~~ $X_S^* \gg R_{sat}$ \  (b)\cr\cr}
\label{eqdif}
\end{equation}

with $R_{sat}\sim (X_S+X_S^*)^{1/3}(FX_S^2\tau_e)^{-1/3}$, which gives for
the crossover between regimes (a) and (b) : $X_S^*(crossover) \sim (FX_S^2\tau_e)^{-1/2}$.

\vspace{1cm}

{\bf direct \ impingement \ growth} $X_S \ll 1$

\begin{equation}
N_{sat}\sim\cases{(F\tau_e)^{2/3} & {\rm if}~~~~$X_S^* \ll 1$ \ (a)\cr\cr
            ({F\tau_e})^{2/3}{X_S^*}^{-2/3}& {\rm if}~~~~$1 \ll X_S^* \ll R_{sat}$ \ (b)\cr\cr
                   F\tau_e & {\rm if}~~~~ $X_S^* \gg R_{sat}$ \  (c)\cr\cr}
\label{eqdir}
\end{equation}

with $R_{sat}\sim ({F\tau_e})^{-1/3} {X_S^*}^{1/3}$, which gives for
the crossover between regimes (a) and (b) : $X_S^*(crossover) \sim (F\tau_e)^{-1/2}$.

\newpage

\begin{center}
{\bf Appendix B}
\end{center}

\vspace{2cm}

I present here the summary of the different limits
of growth of 3d islands in presence of evaporation and/or defects. 
These results are derived in detail in Ref. \cite{pyr} from the
resolution of rate-equations similar to those presented in Appendix A. 
For each regime, I give in the order the saturation
island density $N_{sat}$, the thickness at saturation $e_{sat}$
(i.e. the thickness when the island density first reaches its
saturation value), the thickness at coalescence $e_{c}$ (i.e. the
thickness when the island density starts to decrease due to
island-island coalescence), and the scaling kinetics of the mean island
radius as a function of time before the saturation island density
is reached. I use $l_{CC}=(F \tau)^{1/7}$ for 3d islands \cite{pyr} and
 $X_S=\sqrt{\tau_e/\tau}$. 

\vspace{.7cm}

{\bf Clean substrate (no defects)}

\vspace{.5cm}

high evaporation : $X_S \ll l_{CC} \ll l_{def}$

\hspace{2cm} $N_{sat} \sim [F\tau_e(1+X_s^2)]^{2/3}$

\hspace{2cm} $e_{sat} \sim e_{c} \sim [F\tau_e(1+X_s^2)]^{-1/3}$

\hspace{2cm} $R \sim Ft$

\vspace{.5cm}

low evaporation : $l_{CC} \ll X_S \ll l_{def}$ or $l_{CC} \ll l_{def} \ll X_S$

\hspace{2cm} $N_{sat}\sim\left({F\over D}\right)^{2/7}$   

\hspace{2cm} $e_{sat}\sim e_{c}\sim \left({D\over F}\right)^{1/7}$

\hspace{2cm} $R\sim(FDt^2)^{1/9} \sim t^{2/9}$  

\vspace{1cm}

{\bf Dirty substrate (many defects)}

\vspace{.5cm}

high evaporation : $X_S \ll l_{def} \ll l_{CC}$

\hspace{2cm} $N_{sat}\sim c$  

\hspace{2cm} $e_{sat}\sim {1\over [1+X_s^2]}$

\hspace{2cm} $e_c\sim {1\over c^{1/2}}$

\hspace{2cm} $R\sim Ft$ 

\vspace{.5cm}

low evaporation : $l_{def} \ll X_S \ll l_{CC}$ or $l_{def} \ll l_{CC} \ll X_S$

\hspace{2cm} $N_{sat}\sim c$

\hspace{2cm} $e_{sat}\sim c$

\hspace{2cm} $e_c\sim {1\over c^{1/2}}$

\hspace{2cm} $R\sim \left({Ft\over c}\right)$ for $t \leq c/F$, i.e. before saturation

\hspace{2cm} $R\sim (Ft/c)^{1/3}$  between saturation and coalescence ($c/F \leq t \leq 1/Fc^{1/2}$).

\newpage

\begin{center}
{\bf Table I}

Principal symbols and terms used in this paper. The natural length unit in
the model corresponds to the mean diameter of an incident cluster.
\end{center}

{\it 
\begin{tabbing}
Symbols and terms\hspace{.7cm} \= Units, Remarks \\
\vspace{.3cm}
island \> structure formed on the surface by aggregation of clusters\\
$n$ \>  number of atoms of the cluster\\ 
$d$ \> cluster diameter in nm, $d=d_0 n^{1/3}$ where $d_0$ depends on the element\\ 
site \> area occupied by a cluster on the surface site=$\pi d^2/4$\\
ML \> monolayer : the amount of matter needed to cover uniformly the\\
\hspace{1cm}     \>  substrate with one layer of clusters (1 cluster per site)\\
F \> Impinging flux expressed in monolayers (or clusters per site) 
per second\\ 
$\tau$ \> Diffusion time : mean time needed for a cluster to make a "jump"\\ \hspace{1cm}   \>  between two sites (in seconds)\\
$\tau_e$ \> evaporation time : mean time before a monomer evaporates from the
surface\\
$X_S$ \> mean diffusion length on the substrate before desorption : $X_S=\sqrt{D\tau_e}$\\
$\phi$ \> Normalized flux ($\phi=F\tau$) expressed in clusters per site\\ 
D  \> Diffusion coefficient expressed in $cm^2 s^{-1}$ (D=site/4$\tau$)\\
e  \> mean thickness of the film, $e=Ft$ where $t$ is the deposition time\\
$\theta$ \> coverage; fraction of the substrate covered by the clusters\\
$N_t$ \> island density on the surface, expressed per site\\
$N_{sat}$ \> saturation (maximum) island density on the surface, 
expressed per site\\
$\rho$ \>  density of isolated clusters on the surface, expressed per site\\
$C_{sat}$ \> condensation coefficient (ratio of matter actually present on \\
\hspace{1cm}  \>  the substrate over the thickness) at saturation\\
$\l_{CC}$ \> the island-island distance at saturation when there is no
evaporation\\

\end{tabbing}
}

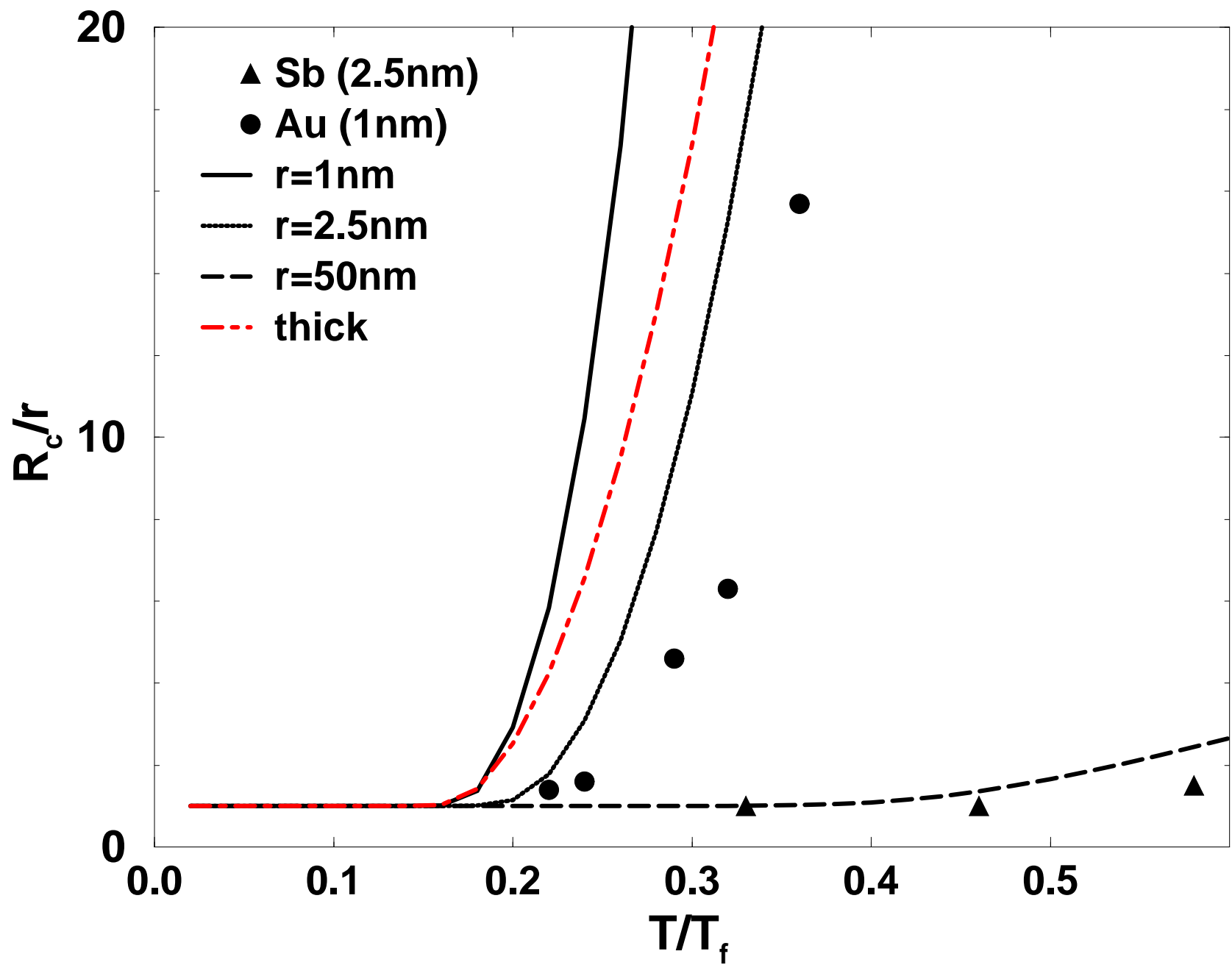
\end{document}